**Generalized Information Ratio**

Zhongzhi (Lawrence) He*

This version: April 2018

## Abstract

Alpha-based performance evaluation may fail to capture correlated residuals due to model errors. This paper proposes using the Generalized Information Ratio (*GIR*) to measure performance under misspecified benchmarks. Motivated by the theoretical link between abnormal returns and residual covariance matrix, *GIR* is derived as alphas scaled by the inverse square root of residual covariance matrix. *GIR* nests alphas and Information Ratio as special cases, depending on the amount of information used in the residual covariance matrix. We show that *GIR* is robust to various degrees of model misspecification and produces stable out-of-sample returns. Incorporating residual correlations leads to substantial gains that alleviate model error concerns of active management.

---

* Goodman School of Business, Brock University, 500 Glenridge Ave., St. Catharines, Ontario, Canada. Email: zhe@brocku.ca.
I am grateful to Ken French who provided data for the Fama-French factors and the portfolio returns. All errors are mine.

## Generalized Information Ratio

Measuring skills of actively managed portfolios has been of long-standing interest to financial economists. In the academic literature on performance evaluation, alpha-based performance measures have evolved from a single factor model to multi-factor benchmarks, the most prominent ones being the Fama and French's (1993) three-factor model and the Carhart's (1997) four-factor model that adds momentum. With the recent success of the five-factor model of Fama and French (2015, 2016) and the q-factor model of Hou, Xue, and Zhang (2015, 2016), additional investment- and profitability-based factors may be required to measure skills. So long as these new factors capture style effects induced by systematic variations in asset returns, their inclusion to performance evaluation models is necessary because managers should be rewarded for their ability to generate fund-specific abnormal returns (alphas), but not for the fund's exposures to common factors (betas). This implies that the widely used Carhart's four factors might be mean-variance inefficient (Roll, 1978), making performance evaluation an ambiguous undertaking.

The performance evaluation literature has long documented that alphas-based measures are susceptible to benchmark error and other forms of model misspecification.[1] Given the challenge to choose the appropriate benchmarks coupled with managers' incentive to engage in benchmark gaming, properly measuring skills that are robust to the choice of benchmarks is of enormous relevance for the well-being of investors.[2] This paper proposes a ratio-based performance measure robust to benchmark error. The central theme of the paper is that the covariance matrix of residual returns (left unexplained by the benchmark model) contains important information in measuring skills. Making full use of this information leads to substantial gains from the Generalized Information Ratio (*GIR* hereafter) that adjusts alphas by residual risk. The measure is so named because it generalizes the single-asset measure of Information Ratio ($IR$ hereafter) to the investment pool of multiple assets in order to fully exploit the residual covariance information.[3] $GIR$ is theoretically motivated by the insightful work of MacKinlay (1995) and MacKinlay and Pastor (2000) who demonstrate that an underlying link between abnormal returns $\alpha$ and the residual covariance matrix $\Sigma$ must exist to prevent arbitrage. They show that in the absence of such a link, unrealistic risk-return trade-off positions can be created with a small number of assets. The link can be further justified by Berk and Green (2004) on the endogenous relation between fund performance and flows. In their model, true skills must be learned from noisy active returns. The absence of the link (e.g., positive alpha with close-to-zero variance) implies that skilled managers are easily identifiable so their funds attract disproportionate amount of inflows. The decreasing returns to scale will erode superior performance; Pastor and Stambaugh (2015) provide pervasive evidence of decreasing returns to scale at fund level. Using a novel method of optimal transport mapping, we derive $GIR$ as alphas scaled by the inverse square root of the residual covariance matrix, i.e., $GIR = \Sigma^{-1/2}\alpha.$ To convert the covariance information into adjusted alphas, the scaling matrix $\Sigma^{-1/2}$ plays two roles: 1) it de-correlates residual returns so that adjusted alphas truly measure fund-specific skills without any correlation effects; 2) it standardizes residual variances to unity so that the power of performance evaluation (Kothari and Warner, 2001) is incorporated into the measure. It is straightforward to show that $GIR$ nests $\alpha$ and $IR$ as special cases, depending the amount of information used in the residual covariance matrix to construct the performance measure. While $GIR$ makes full use of covariance information embedded in $\Sigma$, $IR$ uses only information in the diagonal elements of $\Sigma$, and $\alpha$ uses none of residual information.

[1] For alphas-based measures, Roll (1978) argues that the use of inefficient benchmarks can result in arbitrary ranking of investment managers. Grinblatt and Titman (1994) show that the choice of benchmarks matters much more for performance inferences than the choice of measures. On the other hand, if fund holding data are available, holdings-based measures (e.g., Grinblatt and Titman, 1989; Daniel, Grinblatt, Titman, and Wermers, 1997) are more robust to benchmark error. This paper does not consider holdings-based measures.

[2] Sensoy (2009) documents a strategic fund behavior to self-designate mismatched benchmarks to attract inflows. He reports that nearly one-third of diversified US equity funds specify a benchmark that does not match their actual style. In a similar vein, Elton, Gruber, and Blake (2003) argue that fund managers might deviate from their investment universe in order to beat their declared benchmark. They find that funds have significant exposure to size and value or growth benchmarks that are not captured by their stated benchmark. In light of the difficulty to choose benchmarks, Wermers (2011) suggests "Because no model is perfect, however, a researcher should attempt to apply as many models as is practical, reasonably adding and changing assumptions about benchmarks and model specifications to test for the sensitivity of measured performance to different benchmarks and models." (p. 540)

To see why $GIR = \Sigma^{-1/2}\alpha$ so defined is robust to benchmark error, consider that a group of style-oriented managers employ prior returns as the main signal to pick stocks in the funds they actively manage.[4] Unaware of the underlying return-generating process, an econometrician who uses the Fama-French three factors as the passive benchmarks is likely to find that the group of funds exhibit large alphas in magnitude. That is, using alphas as the performance measure under a misspecified model may bias the assessment of manager ability. However, still under the same misspecified model, if the econometrician looks beyond alphas to examine the structure of the covariance matrix of residual returns, he will discover a systematic pattern: not only large alphas (in magnitude) are accompanied by large residual variances, but the cross products of alphas are also positively correlated with the covariance values. This is because the portion of returns unexplained by the benchmarks is left as active returns defined as $r_t = \alpha + \varepsilon_t$, where $\alpha$ is a vector of true skills and $\varepsilon_t$ collects the noise term for each fund. By MacKinlay's (1995) arbitrage argument and Berk and Green's (2004) theoretical model, the size of $\alpha$ must be positively related to $\Sigma = cov(\varepsilon)$ to prevent fund size from growing excessively to wipe out abnormal returns. As such, our new performance measure $GIR = \Sigma^{-1/2}\alpha$ essentially scales down large sizes of alphas by large standard deviations (i.e., standardization), and it also adjusts the cross-products of alphas by the corresponding off-diagonal values in the residual covariance matrix (i.e., decorrelation), making the *GIR* estimates closer to the true values than using other measures that do not account for residual correlations. We derive these properties in theory, demonstrate an illustrative example, and present large-scale simulation results.

---

[3] Information ratio ($IR$) or appraisal ratio was originally proposed by Treynor and Black (1973). Calculated as the ratio of alpha to standard error of residuals, $IR = \alpha/\sigma_e$ has become the most widely used performance measure to gauge the performance of fund managers in active asset management (Goodwin, 1998; Grinold, 1992, 1995).

[4] This example is motivated by Grinbaltt, Titman, and Wermers (1995) who find that the majority of mutual fund managers use momentum as part of their stock picking strategies. Similarly, Carhart (1997) argues that some apparent persistence in fund performance is due simply to momentum in stock returns.

Benchmark error is not the only source of correlated residuals that should be controlled in order to measure true skills. The performance evaluation literature has documented numerous other sources of commonalities in residual returns. Wermers (2011) summarizes these commonalities in terms of covariates of skilled managers, including social connections (e.g., Cohen, Frazzini, and Malloy, 2008), academic background (e.g., Chevalier and Ellison, 1999; Gottsman and Morey, 2006), and coinvestment (e.g., Jones and Shanken, 2005; Cohen, Coval, and Pastor, 2005), among others. Extant studies using alphas-based measures attempt to construct empirical proxies to capture these correlated residuals. For example, Pastor and Stambaugh (2002) suggest including non-benchmark assets to improve inferences about manager skills. Hunter, Kandel, Kandel, and Wermers (2014) construct an active peer benchmark to account for commonalities in residual returns. These proxies have demonstrated empirical success in their respective design of manager covariates. However, it is conceivable of certain scenarios that these proxies may not capture systematic residual correlations. Specifically, the active peer benchmark may not capture commonalities *across* peer groups (Hunter, Kandel, Kandel, and Wermers, 2014, p. 9), and the choice of non-benchmark assets may be somewhat arbitrary (Pastor and Stambaugh, 2002, p. 325). Take another example as motivated by Gottsman and Morey (2006), a subgroup of skilled managers graduated from an elite MBA program may share private information across various peer groups with different style orientations. Such an unobserved factor seems to be unrelated to the within-group peer benchmark or any of the non-benchmark assets, nor is it relevant to the choice of passive benchmarks. Nonetheless, the effect of this omitted factor on fund performance will be left in active returns $\alpha + \varepsilon_t$, and is likely to leave a trace as a positive $\alpha$-$\Sigma$ link in the subgroup of the fund evaluation pool. Such a link is expected to exist, otherwise the diseconomics of scale will immediately erode the superior performance (Berk and Green, 2004). Thus, using *GIR* as the performance measure under a carefully chosen model, the alphas of this subgroup of funds will be proportionally adjusted by the scaling matrix, mitigating the correlated effect for this subgroup of skilled managers.

In sum, given that benchmark error or gaming can take any form (i.e., no model is perfect) and there are uncountable ways that manager skills may co-vary with each other (documented or not), it is practically implausible to identify all sources of commonalities in residual returns and account for them. This paper proposes $GIR = \Sigma^{-1/2}\alpha$ as a robust performance measure to account for correlated residuals without identifying their sources.[5] Both benchmark error and manager covariates are treated as omitted factors of a misspecified model, which is compared to the true model for competing performance measures. The theory part of the paper formally derives *GIR* and shows the decorrelation and standardization roles of the scaling matrix. In addition, a novel distance metric is developed to measure the cross-sectional dispersion of adjusted alphas under the misspecified model relative to that under the true model for a given performance measure.

The empirical part of the paper then examines the potential gains of using *GIR* as compared to alphas and *IR* under differential degrees of model misspecification. Using a wide array of Fama-French characteristics-sorted portfolios to represent investment styles, our simulation procedure is in the spirit of Kothari and Warner (2001); however, we extend their single-portfolio design to multiple portfolios randomly selected at each period. Motivated by Grinblatt and Titman (1994) who compare the choice of benchmarks with the choice of performance measures, we introduce the notion of substitution effect; namely, a robust performance measure under a poor model can substitute the effect of alternative measures under a less misspecified model. To illustrate, for the case of $n$ = 25 funds and $L$ = 120 months of evaluation period, using *GIR* under *CAPM* (i.e., the most misspecified model) produces the same cross-sectional dispersion as using alphas under the less misspecified Fama-French three-factor model. Furthermore, the substitution effect becomes stronger as the fund pool gets large. For example, for the case of $n$ = 200 funds and $L$ = 360 months, using *GIR* to measure performance under *CAPM* produces an even lower dispersion than using *IR* under the Fama-French five-factor model (i.e., the least misspecified model). In addition, we also examine the out-of-sample performance of using different measures to rank funds then form long-short portfolios. We find that *GIR* produces the most stable out-of-sample returns accompanied by higher *t*-statistics, and the result is the strongest for large number of funds under the most misspecified model. For example, alpha standard deviation using the *GIR* ranking is typically one-third or less of that using the *IR* or alpha ranking for $n$ = 100 or 200 funds under *CAPM*. Overall, these results contribute to our understanding of performance evaluation under misspecified models as summarized below. First, the more misspecified the model is, the greater are the potential gains of using *GIR*, shown both in theory and by simulation results. Second, decorrelation (i.e., adjusting correlated residuals by the scaling matrix) plays the most important role for *GIR* as a robust performance measure; the larger the number of funds under evaluation, the stronger the decorrelation effect. However, the large number of funds must be accompanied by about twice as long fund return history. We show this stringent requirement using weekly data in the context of accurate estimation of high-dimensional covariance matrix. In short, using higher frequency returns (if available) can be highly effective for the proper use of *GIR* even for a short fund history.

[5] This is in the spirit of "learning across funds" of Jones and Shanken (2005). Within a Bayesian framework, they examine different degrees of learning priors on fund performance and propose a shrinkage estimator of alphas. In comparison, *GIR* does not depend on any prior specifications. Learning across funds is achieved by incorporating residual covariance information into adjusted alphas.

The rest of the paper is organized as follows. Section 1 presents the theoretical link of alphas and the residual covariance matrix and the classic results of the optimal transport theory, based on which *GIR* and the distance metric for measure comparison are derived. Section 2 provides an illustrative example to develop the intuition. Section 3 presents the data and simulation design, and show all the simulation results. Section 4 concludes the paper with future research, and provides guidelines for the proper use of *GIR*.

## 1. Theoretical Foundation

### 1.1 The Link between Alphas and Residual Covariance Matrix

The exposition of this section is based on MacKinlay and Pastor (2000). Suppose an econometrician evaluates the performance of $n$ actively managed portfolios, using a multifactor asset-pricing model consisting of $k$ passive benchmarks. The econometrician has access to $L$ periods of portfolio excess returns (in excess of the risk free rate) and benchmark returns. Portfolio performance is evaluated by the following multivariate time-series regressions:

$$R_t = \alpha + \beta_B B_t + \varepsilon_t \quad (1.1)$$

$$E[\varepsilon_t] = 0,\ E[\varepsilon_t \varepsilon_t'] = \Sigma,\ cov(B_t, \varepsilon_t) = 0 \quad (1.2)$$

where $R_t$ denotes the vector of excess returns on the set of $n$ portfolios in period $t$; $\alpha$ is the $n$-vector of abnormal returns with respect to $k$-benchmark $B$; $\beta_B$ is the $n\times k$ matrix of benchmark exposures; $B_t$ is the $k$-vector of benchmark returns in period $t$; $\varepsilon_t$ is the $n$-vector of residual returns left unexplained by the benchmarks, and the $\Sigma$ is the $n\times n$ residual covariance matrix.

In performance evaluation of actively managed portfolios, each element of the vector of abnormal returns $\alpha$ reflects managerial skills of a given fund. However, the set of passive benchmarks may not be mean-variance efficient to capture all common variations in fund returns (Roll, 1978). In addition, there may exist missing factors due to common strategies among funds, common talents among managers, private information among subgroups, among others, leaving unobserved commonalities in the residuals $\varepsilon_t$. Eq. (1) is thus referred to as the misspecified model. To illustrate the impact of the model misspecification and establish the theoretical link between $\alpha$ and $\Sigma$, consider adding an unobserved orthogonal factor $F$ to eq. (1), such that

$$R_t = a + \beta_B B_t + \beta_F F_t + u_t \tag{2.1}$$

$$E[u_t] = 0,\ E[u_t u_t'] = \Phi,\ cov[B_t, u_t] = 0,\ cov[F_t, u_t] = 0,\ cov[B_t, F_t] = 0 \tag{2.2}$$

$$E[F_t] = \mu_F,\ var(F_t) = \sigma_F^2 \tag{2.3}$$

where $a$ is the $n$-vector of abnormal returns with respect to the $k$-benchmark $B$ and the omitted factor $F$, and $\beta_F$ is the $n$-vector of funds' exposures to the omitted factor $F$ whose return at time $t$ is $F_t$. $\beta_B$ in eq. (2.1) is not altered because $F_t$ is made orthogonal to $B_t$. Eq. (2) is referred to as the true model with well-specified factors ($B$ and $F$) to define the true alphas $a$ and the true residual covariance matrix $\Phi$.

The link between abnormal returns and residual covariance matrix can be seen by comparing eqs. (1.1) and (2.1), yielding

$$\alpha = a + \beta_F \mu_F \tag{3}$$

$$\Sigma = \beta_F \beta_F' \sigma_F^2 + \Phi = (\alpha - a)(\alpha - a)'/S_F^2 + \Phi \tag{4}$$

where $S_F \equiv \mu_F/\sigma_F$ denotes the Sharpe ratio of the omitted factor $F$. Eq. (3) shows that the unobserved factor $F$ provides an incremental explanatory power of abnormal returns by the amount of $\beta_F \mu_F$. If $F$ explains all the abnormal returns so that the intercept in eq. (2.1) vanishes ($a = 0$), then $F$ becomes the optimal orthogonal portfolio in MacKinlay (1995) and MacKinlay and Pastor (2000). This paper considers the more general and realistic case where the missing common factor $F$ does not fully explain all abnormal returns, so $a \neq 0$.

Eq. (4) establishes the theoretical link between abnormal returns and residual covariance through the part of abnormal returns explained by the omitted factor *F*. Thus, the vector of abnormal returns in eq. (1.1) is embedded in the residual covariance matrix, and such a link can be explicitly exploited in the construction of *GIR*. In the context of performance evaluation under a misspecified model, eq. (4) says that the residual covariance matrix $\Sigma$ contains information about a systematic portion of abnormal returns attributable to some omitted common factors. MacKinlay (1995) provides an economic intuition of the link for asset pricing. Abnormal returns must be accompanied by a common component in the residual covariance matrix to prevent arbitrage opportunities from being realized by an active portfolio with non-zero alpha but close-to-zero residual variance. Pastor and Stambaugh (2000, 2002) incorporate this link into their prior specifications of model parameters. For research on mutual funds, the theoretical model of Berk and Green (2004) provides further economic justification. Manager ability $\alpha$ must be positively correlated with the uncertainty matrix $\Sigma$; otherwise fund skills can be known without learning, causing explosive flows to wipe out abnormal returns.[6]

The theoretical $\alpha$-$\Sigma$ link in eq. (4) implies the following empirical observation: large magnitude and cross-products of $\alpha$ under the misspecified model should be associated with corresponding large variances and covariances in $\Sigma$, relative to their true values $a$ and $\Phi$ under the true model. In performance evaluation of mutual funds, this leads to an empirical test of correlated residuals induced by benchmark error or manager covariates. To illustrate, suppose two skilled managers $i$ and $j$ who share private information produce large alphas $\alpha_i$ and $\alpha_j$ under a misspecified model that omits this common factor. Eq. (4) implies that large $\alpha_i$ and $\alpha_j$ are associated with large $\sigma_i$ and $\sigma_j$, relative to their true values. Manager covariates imply a positive correlation coefficient $\rho_{ij} > 0$. As a result, the covariance term $\sigma_{ij} = \rho_{ij}\sigma_i\sigma_j$ should also be large and proportional to the product of the two alphas $\alpha_i\alpha_j$. When information embedded in correlated residuals is translated into adjusted alphas (*GIR*), large alphas and their cross-products will be correspondingly scaled by large residual covariance and covariance values. Essentially, the proportional scaling tends to diminish the effect of commonalities in residual returns, making adjusted alphas closer to their true values. Whether or not and to what extent of we can benefit from using *GIR* is subject to empirical examination.

[6] In eq. (30) of Berk and Green's model, the absence of the link implies the possibility of $\omega \to \infty$ for young funds ($t$ is small) with large $r_t > 0$, causing fund size $q_t$ to explode.

Empirical work requires an assumption about the structure of $\Phi$ in eq. (4). MacKinlay and Pastor (2000) impose $\Phi$ to be proportional to the identity matrix, i.e., $\Phi = \sigma^2 I$. The identity matrix plays a key role for establishing the $\alpha$-$\Sigma$ link in their paper. They employ the maximum likelihood procedures by restricting $\Phi = \sigma^2 I$ in eq. (4) ($a$ = 0 in their paper), with $\alpha, S_F^2$ and $\sigma^2$ as the free parameters to be estimated. There are no closed-form solutions for their maximum-likelihood estimates, so numerical procedure has to be adopted. Further, they noted that treating $S_F^2$ as a free parameter causes an estimation instability (p. 902); instead, they estimated $\alpha$ conditioning on some carefully specified values of $S_F^2$.

This paper adopts a different estimation methodology. Instead of estimating alphas by imposing the $\alpha$-$\Sigma$ link and $\Phi$ proportional to the identity matrix as the maximum-likelihood restrictions, we convert the covariance information into adjusted alphas based on optimal transport mapping. Specifically, we respectively map the cross-sectional distributions of active returns generated from eqs. (1) and (2) to the target distribution with an identity covariance matrix. This is the classic optimal transport problem, whose solution is known as the optimal transport plan given by a unique optimal mapping matrix *T*, which has a simple form as the inverse square root of the residual covariance matrix under the standard assumption of Gaussian distributions of active returns ($T = \Sigma^{-1/2}$). As a result, the mean of the mapped distribution is alphas scaled by $T$ (i.e., $\Sigma^{-1/2}\alpha$ for the misspecified model and $\Phi^{-1/2}a$ for the true model), which essentially translates residual covariance information into adjusted alphas. Below we present the optimal transport theory along with distance-based metrics, based on which *GIR* is formally derived.

### 1.2 Distance-based Metrics and Optimal Transport Mapping

The distance metrics are based on the optimal transport theory deeply rooted in mathematics (Villani, 2003, 2009) with rich applications in economics (Galichon, 2016) and econometrics (Galichon, 2017). The classic problem (Monge, 1781) is to find the shortest distance or the minimum cost to move the mass of one probability distribution to another. Following Villani (2009, Definition 6.1), we give the following definition of the Wasserstein distance between two probability distributions.[7]

**Definition**: Let ($S$, $d$) be a Polish metric space. Assume that two probability measures $P$ and $Q$ on $S$ are continuous and have finite moments of order $z \in [1, \infty)$. The Wasserstein distance between $P$ and $Q$ is defined by

[7] The Wasserstein distance is also known as the Monge-Kantorovich distance in the optimal transport literature (Villani 2009, Chapter 7). See also Villani (2009, Chapter 6) for a chronological review of this terminology.

$$WD_z(P,Q) = \left[inf \int_S \; d^z(x,y)d\pi(x,y)\right]^{\frac{1}{z}}$$

$$= \inf\{[Ed(X,Y)^z]^{\frac{1}{z}}, \;\; law(X) = P \; and \; law(Y) = Q\} \quad (5)$$

The infimum is taken over all $\pi(x, y)$ in $\Pi(P, Q)$, which is the set of joint probability measures on random variables $X \times Y$ with marginals $P$ on $X$ and $Q$ on $Y$.

Specifically, we are interested in $z = 2$ in this paper, and define the quadratic Wasserstein distance as:

$$WD_2(P,Q) = \inf(E_\pi||X-Y||^2)^{1/2} \quad (6)$$

The infimum is taken over all the transport plan $\pi(x, y)$ in $\Pi(P, Q)$, with the marginal distribution of $P$ on $X$ and $Q$ on $Y$.

Given the definition of $WD_2$, the optimal transport literature proves the following theoretical results.

1) There exists a unique solution to the optimal transport problem of moving the mass of distribution $P$ to that of $Q$. The one-to-one mapping is known as the optimal transport plan $y = T(x)$, where $x \sim P$ is mapped to $y \sim Q$ via $T(x)$.

2) Under the optimal transport plan, random vectors $X \sim P$ and $Y \sim Q$ are maximally correlated with each other.

Above are standard results in the optimal transport theory (Villani, 2003, 2009).[8] $WD_2$ has the economic interpretation as the minimum expected cost to transport the mass of distribution $P$ to $Q$. In general, there exists no closed-form formula for $WD_2$ or $T(x)$ for general distributions. Fortunately, when $P$ and $Q$ are Gaussian, closed-form formula for $WD_2$ and $T(x)$ can be derived, with the key results summarized in the following theorem.

**Optimal Transport Theorem for Gaussian Distributions**

1) Let $P$ and $Q$ be two Gaussian measures on $R^n$ with finite second moments such that $P{\sim}N(\alpha_p, \Sigma_p)$ and $Q{\sim}N(\alpha_q, \Sigma_q)$, where $\alpha_p$ and $\alpha_q$ are two $n{\times}1$ vectors of mean, $\Sigma_p$ and $\Sigma_q$ are two $n{\times}n$ symmetric, positive-definite covariance matrices, respectively.

   The quadratic Wasserstein distance ($WD_2$) between $P$ and $Q$ is given by

$$WD_2 = \sqrt{||\alpha_p - \alpha_q||^2 + ||\Sigma_p - \Sigma_q||} \quad (7.1)$$

[8] See Villani (2003, Theorem 2.12) for the proof of the existence and uniqueness of the solution in 1), with the maximum correlation property in 2) as a corollary. See also Galichon (2016, 2017) for the applications in economics and econometrics.

and

$$||\Sigma_p - \Sigma_q|| = tr(\Sigma_p + \Sigma_q - 2(\Sigma_p^{1/2}\Sigma_q\Sigma_p^{1/2})^{1/2}) \quad (7.2)$$

where $||\alpha_p - \alpha_q||$ is the Euclidean 2-norm of the mean difference vector; $||\Sigma_p - \Sigma_q||$ denotes the distance between the two covariance matrices; $Tr(\cdot)$ is the trace operator of a matrix; $\Sigma_p^{1/2}$ is the square root of the covariance matrix $\Sigma_p$ such that $\Sigma_p = \Sigma_p^{1/2}\Sigma_p^{1/2}$. For symmetric and positive-definite covariance matrix $\Sigma_p$, its ($n \times n$) square root matrix $\Sigma_p^{1/2}$ is unique, symmetric, and positive-definite.[9]

2) There exists an optimal transport plan that uniquely maps random variable $r_p \sim P$ to $T(r_p) \sim Q$ via

$$T(r_p) = T_p r_p \quad (8)$$

where the $n \times n$ symmetric, positive-definite optimal mapping matrix is given by

$$T_p = \Sigma_p^{-1/2}(\Sigma_p^{1/2}\Sigma_q\Sigma_p^{1/2})^{1/2}\Sigma_p^{-1/2} \quad (9)$$

3) The optimal transport plan has the following properties:
   a) $T_p\Sigma_p T_p = \Sigma_q$ (10)
   b) It has the shortest distance or the minimal expected cost to move the mass of *P* to *Q*; and
   c) *P* and *Q* are maximally correlated with each other.

The Appendix outlines the proof with technical details. Based on the above classic results, *GIR* and a distance metric for measure comparison are derived below.

### 1.3 Generalized Information Ratio (*GIR*) and Empirical Hypotheses

Active returns produced by the misspecified model eq. (1) are defined as $r_{p,t} \equiv R_t - \beta_B B_t = \alpha + \varepsilon_t$, whose conditional distribution is assumed to be multivariate normal, serially uncorrelated and homoscedastic over time, i.e., $r_{p,t}|X_{p,t} \sim N(\alpha, \Sigma)$, where $X_{p,t} = [R_t, B_t]$ is the set of conditioning variables of the misspecified model.[10] Subscript "*p*" indicates the misspecified model as model $p$.

To apply the optimal transport mapping for Gaussian distributions as in eqs. (8) to (10), let *P* be $r_{p,t}|X_{p,t} \sim N(\alpha, \Sigma)$ and define its target distribution $P^*$ to be $r^*_{p,t}|X_{p,t} \sim N(E[r^*_{p,t}], I_n)$, where $I_n$ denotes the

---

[9] The symmetric and positive-definite $\Sigma_p^{1/2}$ is computed using the Schur algorithm (Deadman, Higham, and Ralha, 2013). Python library scipy.linalg.sqrtm() implements this algorithm.

[10] For asset *i*, its active return $r_{i,t}$ can be thought of period *t*'s return on a benchmark-neutral fund by taking one unit of long position on asset $R_{i,t}$, and $\beta_i$ units of short positions on *K* tradable benchmarks. Active returns and residual returns are used interchangeably in this paper.

$n \times n$ identity matrix. Now consider a mapping from $P$ to $P^*$ ($T$: $P \rightarrow P^*$) via the optimal transport plan $r_{p,t}^* = T(r_{p,t}) = T_p r_{p,t}$ such that

$$r_{p,t}^* | X_{p,t} \sim N\left(T_p E\left[r_{p,t}\right], T_p \Sigma T_p\right) = N(T_p \alpha, I_n) \quad (11)$$

When the target covariance is the identity matrix $I_n$, the optimal mapping matrix of eq. (9) simplifies to

$$T_p = \Sigma^{-\frac{1}{2}} (\Sigma^{1/2} I_n \Sigma^{1/2})^{1/2} \Sigma^{-\frac{1}{2}} = \Sigma^{-\frac{1}{2}} \quad (12)$$

which becomes the inverse square root of the residual covariance matrix of the misspecified model. As seen, the scaling matrix $\Sigma^{-1/2}$ de-correlates residual returns and standardizes residual variances to unity to make an identity residual covariance matrix $I_n$. Consequently, the mean of the mapped distribution in eq. (11) gives the definition of the $n$-vector *GIR* for the misspecified model:

$$GIR_p = \Sigma^{-\frac{1}{2}} \alpha \quad (13.1)$$

That is, $GIR_p$ is alphas scaled by the inverse square root of the residual covariance matrix of the misspecified model. $GIR_p$ links $\alpha$ through both variance and covariance information in $\Sigma$, so $GIR_p$ is interpreted as abnormal returns per unit of *covariance* risk of residual returns.

$GIR_p$ reduces to $IR_p$ when the residual covariance matrix $\Sigma$ is assumed to be a diagonal matrix $D_p$, i.e.,

$$IR_p = D_p^{-\frac{1}{2}} \alpha \quad (13.2)$$

For a given asset $i$, $IR_{p,\mathrm{i}} = \alpha_i / \sigma_{p,i}$, where $\sigma_{p,i}$ is the square root of its diagonal element in $\Sigma$. $IR_p$ links $\alpha$ through only the variance elements but ignores the covariance information embedded in the off-diagonal components in $\Sigma$, so $IR_p$ is interpreted as abnormal returns per unit of *heteroskedastic* risk of residual returns.

Finally, if we further assume that all diagonal elements of $D_p$ are identical, then $IR_p$ reduces to alphas normalized to unity risk, i.e.,

$$\alpha_p^* = \alpha / \sigma_{p,e} \quad (13.3)$$

where $\sigma_{p,e}$ is a scaler, set equal to the average of the square root of diagonal elements of $\Sigma$, i.e., $\sigma_{p,e} = \sum_{i=1}^{n} \sigma_{p,i} / n$. $\alpha_p^*$ is thus interpreted as abnormal returns per unit of *homoscedastic* risk of residual returns. $\alpha_p^*$ does not use any information contained in the residual covariance matrix, and it produces identical performance rankings as alpha, which is obtained by setting $\sigma_{p,e} = 1$. $\alpha_p{}^*$ is made to have the same scale as $IR_p$ and $GIR_p$ for comparison purpose.[11]

Similarly, active returns produced by the true model eq. (2) are defined as $r_{q,t} \equiv R_t - \beta_B B_t - \beta_F F_t = a + u_t$, whose conditional distribution is multivariate normal, serially uncorrelated and homoscedastic over time, i.e., $r_{q,t}|X_{q,t} \sim N(a, \Phi)$, where $X_{q,t} = [R_t, B_t, F_t]$ is the set of conditioning variables of the true model. Subscript "$q$" indicates the true model as model $q$. Let $Q$ be $r_{q,t}|X_{q,t} \sim N(a, \Phi)$ and define its target distribution $Q^*$ to be $r^*_{q,t}|X_{q,t} \sim N(E[r^*_{q,t}], I_n)$. Consider a mapping from $Q$ to $Q^*$ ($T$: $Q \to Q^*$) via the optimal transport plan $r^*_{q,t} = T(r_{q,t}) = T_q r_{q,t}$ such that

$$r^*_{q,t}|X_{q,t} \sim N\left(T_q E\left[r_{q,t}\right], T_q \Phi T_q\right) = N(T_q a, I_n) \quad (14)$$

where the optimal mapping matrix of eq. (9) simplifies to

$$T_q = \Phi^{-\frac{1}{2}}(\Phi^{1/2} I_n \Phi^{1/2})^{1/2} \Phi^{-\frac{1}{2}} = \Phi^{-\frac{1}{2}} \quad (15)$$

which becomes the inverse square root of the residual covariance matrix of the true model. We can therefore define $GIR_q$, $IR_q$, and $\alpha_q{}^*$ for the true model as:

$$GIR_q = \Phi^{-\frac{1}{2}} a \quad (16.1)$$

$$IR_q = D_q^{-\frac{1}{2}} a \quad (16.2)$$

$$\alpha^*_q = a / \sigma_{q,e} \quad (16.3)$$

where $D_q$ is the $n$-vector diagonal matrix whose values are taken from the diagonal elements of $\Phi$. $\sigma_{q,e}$ is a scaler set equal to the average of the square root of diagonal elements of $\Phi$, i.e., $\sigma_{q,e} = \sum_{i=1}^{n} \sigma_{q,i} / n$.

In sum, we have derived *GIR* for both the misspecified model in eq. (13.1) and the true model in eq. (16.1), nesting *IR* and $\alpha^*$ as special cases depending on the amount of information used in the covariance matrix. In other words, the three measures are abnormal returns per unit of residual risk assumed with different information structure. *GIR* makes full use of covariance information. *IR* assumes a heteroskedastic residual structure that only uses information in the diagonal elements of the residual covariance matrix. And $\alpha^*$ assumes a homoscedastic structure using no residual information. The main purpose of the paper is to evaluate the three performance measures under differential degrees of model misspecification against those under the true model. This is accomplished by measuring the distance between the distribution of active returns generated from the misspecified model and that generated from the true model. The distance metric is theoretically derived by eq. (7). This metric is optimal in the sense of the shortest distance or the minimum cost to transport the mass of $P^*$ in eq. (11) to $Q^*$ in eq. (14), or the maximum correlation between $P^*$ and $Q^*$. Note that both $P^*$ and $Q^*$ have identical variance of an identity matrix, hence the distance metric given by eq. (7) is solely determined by the Euclidean 2-norm of the difference between $GIR_p$ in eq. (13.1) and $GIR_q$ in eq. (16.1). Because the 2-norm is the square root of the sum of squared elements in the vector, the distance metric is basically the square root of sum squared difference between $GIR_p$ and $GIR_q$. The distance between $IR_p$ and $IR_q$, and that between $\alpha_p{}^*$ and $\alpha_q{}^*$ are just two special cases of the distance between $GIR_p$ and $GIR_q$. As such, the distance metrics for the three performance measures are expressed below:

$$||GIR_p - GIR_q|| = ||\Sigma^{-\frac{1}{2}}\alpha - \Phi^{-\frac{1}{2}}a|| \tag{17.1}$$

$$||IR_p - IR_q|| = \sqrt{\sum_{i=1}^{n}(\frac{\alpha_i}{\sigma_{p,i}} - \frac{a_i}{\sigma_{q,i}})^2} \tag{17.2}$$

$$||\alpha_p^* - \alpha_q^*|| = \sqrt{\sum_{i=1}^{n}(\frac{\alpha_i}{\sigma_{p,e}} - \frac{a_i}{\sigma_{q,e}})^2} \tag{17.3}$$

[11] *GIR*, *IR* and $\alpha^*$ are all ratio-based measures that adjust for model power. Alpha-based measures can be easily obtained by multiplying each ratio-based measure by the same scaling factor $\sigma_e$, the average standard error of residuals. By doing so, alpha-based measures are comparable under the same model, but not *across* models because of unequal power (i.e., different models produce different $\sigma_e$). For this reason, we use ratio-based *GIR*, *IR* and $\alpha^*$ for comparison purpose. In practical applications, ratio-based measures have an added benefit of invariance to leverage, a property not present in alpha-based measures subject to the leverage bias (Modigliani and Pogue, 1974).

The distance metrics so defined measure the cross-sectional dispersion of adjusted alphas under the misspecified model relative to that under the true model. The performance measure with the shortest distance between the misspecified model (model $p$) and the true model (model $q$) is deemed the most reliable one. It means that when a poor model is chosen for performance evaluation, the measure with the lowest dispersion (relative to its true value) is the least sensitive or the most robust to model misspecification. Given the theoretical link of $\alpha$ and $\Sigma$ described in Section 1.1, empirical predictions for comparing the three performance measures (*GIR*, *IR* and $\alpha^*$) are formulated below.

- <u>*Hypothesis 1*</u>:

  The theoretical link of $\alpha$ and $\Sigma$ implies $||GIR_p - GIR_q|| < ||IR_p - IR_q|| < ||\alpha_p^* - \alpha_q^*||$. Specifically,

  a) If both residual variances and residual covariances contain useful information in measuring skills, then *GIR* should produce the shortest distance;

  b) If only information in residual variances is useful (i.e., uncorrelated residuals but heteroskedastic residual variances), then *IR* should produce the shortest distance;

c) If none of residual information is useful (i.e., uncorrelated residuals and homoscedastic residual variance), then $\alpha^*$ should produce the shortest distance.

- *Hypothesis 2*:

  Under *Hypothesis 1*, the stronger the explanatory power of the omitted factor, the greater the gains of using the respective measure as compared to the other two measures.

*Hypothesis 1* forms the empirical predictions for comparing the three performance measures. *1.a* is implied from the theoretical link. If the residual covariance matrix does not contain much useful information, or if the noise due to sampling error outweighs the information content, then *1.b* or *1.c* should be observed. *Hypothesis 2* stipulates the effect of differential degrees of model misspecification on performance evaluation. In the special (but unrealistic) case where the missing common factor $F$ explains all abnormal returns, i.e., $a = 0$ in eq. (2.1), the theoretical link exhibits its full merit in the estimation of abnormal returns. To what extent are the potential gains of using the robust measure is subject to empirical examination below.

## 2. An Illustrative Example

Before presenting the large-scale simulation results, we examine the set of five momentum portfolios illustrated earlier. This example is stylized and does not aim to represent real style effects. However, it allows us to view the regression coefficients, the 5×5 residual covariance matrices, and the 5×5 scaling matrix used to construct the *GIR*, *IR*, and $\alpha^*$ performance measures. In particular, it is informative to examine how these three measures produce differential results under the misspecified model and the true model, and why *GIR* produces robust ranking results despite the use of misspecified model. Much of the intuition and illustrative results carry over to the large-scale simulation results in the next section.

Suppose one evaluates the relative performance of five momentum portfolios. Unaware that these portfolios are formed by sorting prior returns, he uses the Fama-French three-factor model (*FF3*) as the benchmark model. However, the true model should be Carhart's (1997) four-factor model (*Carhart4*) with the momentum factor (*UMD*) added to *FF3*.

[Insert Table 1 here]

To examine the impact of the omitted factor (*UMD* in this example), we follow Fama and French (2015) and run a spanning regression of *UMD* on the three factors for the period of 1963:07 – 2016:12, and obtain the following results:

$$UMD_t = \underset{(5.39)}{0.89} - \underset{(-4.79)}{0.19} \times MKT_t + \underset{(0.27)}{0.01} \times SMB_t - \underset{(-5.98)}{0.35} \times HML_t + \varepsilon_t, \qquad R^2 = 0.07$$

Then we define the orthogonal *UMD* (*UMDo*) as the sum of the intercept and residual from the above regression. *UMDo* has a mean of 0.89% per month and monthly standard deviation of 4.07%. Thus, *Carhart4* is *FF3* plus *UMDo*.

Panel A of Table 1 reports the regression coefficients and *t*-statistics for *FF3* on the left side and for *Carhart4* on the right side. If *FF3* is used as the passive benchmarks, the five portfolios exhibit highly significant abnormal returns, ranging from -0.74% ($t$ = -5.53) to 0.47% ($t$ = 6.35) per month. If the true model is used to evaluate performance, abnormal returns range from -0.07% ($t$ = -1.37) to 0.14% ($t$ = 3.86) per month. These results confirm earlier studies (e.g., Grinblatt and Titman, 1994) that benchmark misspecification can have a dramatic effect on alpha-based performance evaluation. Coefficients on *MKT*, *SMB*, and *HML* are identical under *FF3* and *Carhart4* because *UMDo* is made orthogonal to the three factors. Therefore, the additional explanatory power in reducing alphas and in improving $R^2$s is solely attributed to the omitted orthogonal factor *UMDo*.

Panel B of Table 1 presents the residual covariance matrices Σ under *FF3* and Φ under *Carhart4.* As shown in eq. (4), $\alpha$ and Σ are related by $\Sigma = (\alpha - a)(\alpha - a)'/S_F^2 + \Phi$, where $\alpha$ and $a$ are the intercepts shown in Panel A, and $S_F = \mu_F / \sigma_F$ = 0.89%/4.07% = 0.218. Empirical observations of the $\alpha$-Σ link are readily seen. The largest magnitude of alpha difference ($\alpha_1 - a_1$ = -0.74% + 0.07% = -0.67%, relative to its true value) is accompanied by the highest residual variance of $\sigma_1^2$ = 10.86bp; the second largest magnitude of alpha difference ($\alpha_5 - a_5$ = 0.47% – 0.12% = 0.35%) is accompanied by the second highest residual variance of $\sigma_5^2$ = 3.37bp; and the third largest magnitude of alpha difference ($\alpha_2 - a_2$ = -0.16% – 0.14% = -0.30%) is accompanied by the third highest residual variance of $\sigma_2^2$ = 2.68bp. This systematic relation implies that, for a given asset $i$, by dividing $\alpha_i$ by its standard deviation $\sigma_i$, *IR* should produce a less dispersed (relative to the true value) performance measure than $\alpha^*$ that divides $\alpha_i$ by a constant scaler. Intuitively, larger magnitude of alphas is scaled down more heavily than smaller magnitude of alphas. Now examine the relation between the cross-products of ($\alpha$ - $a$) and off-diagonal elements in Σ. We find that the top three residual covariances $\sigma_{15}$ = -4.27bp, $\sigma_{12}$ = 4.03bp, and $\sigma_{25}$ = -2.18bp, are respectively accompanied by the three largest cross-products of alpha differences $(\alpha_1 - a_1)(\alpha_5 - a_5)$ = -0.24bp, $(\alpha_1 - a_1)(\alpha_2 - a_2)$ = 0.20bp, and $(\alpha_2 - a_2)(\alpha_5 - a_5)$ = -0.11bp, with the same sign and in the same order. The information embedded in the residual covariances is explained below. The use of the misspecified model *FF3* tends to produce extreme outperforming portfolio (top winner) and extreme underperforming portfolio (bottom loser), implying that the two managers possess negatively correlated skills ($\rho_{15} < 0$). Given large $\alpha_1$ and $\alpha_5$ in magnitude, the covariance of the two managers ($\sigma_{15} = \rho_{15}\sigma_1\sigma_5$) must also be largely negative and proportional to the product of the alphas. In short, these results empirically support the underlying link between abnormal returns and residual covariance matrix. Therefore, using information embedded in the residual covariance matrix of the misspecified model should improve the estimates of skills relative to the true model. This is demonstrated below by comparing the *GIR*, *IR*, and $\alpha^*$ measures to evaluate the performance of the five momentum portfolios.

Panel B also shows the scaling matrices, $\Sigma^{-1/2}$ under *FF3* and $\Phi^{-1/2}$ under *Carhart4*. One can verify that the two matrices map the corresponding residual covariance matrix to the identity matrix, i.e., $\Sigma^{-1/2}\Sigma\Sigma^{-1/2} = I_n$ and $\Phi^{-1/2}\Phi\Phi^{-1/2} = I_n$. That is, $\Sigma^{-1/2}$ and $\Phi^{-1/2}$ de-correlate residual returns and standardize residual variances. By doing so, information embedded in $\Sigma$ and $\Phi$ is translated into alphas adjusted by the scaling matrix, i.e., $GIR_3 = \Sigma^{-1/2}\alpha$ and $GIR_4 = \Phi^{-1/2}\alpha$. If residual information (due to the omitted *UMD* factor) is indeed useful in measuring skills, then adjusted alphas $\Sigma^{-1/2}\alpha$ should be closer to their true values $\Phi^{-1/2}\alpha$, making *GIR* a more robust performance measure. For example, let us examine *GIR* of the bottom loser portfolio under *FF3*, denoted by $GIR_{3,1}$ = 0.44×(-0.74) + (-0.26)×(-0.16) + 0.04×0.00 + 0.13×0.17 + 0.13×0.47 = -0.20. As seen, information embedded in the covariance of the bottom loser with other portfolios is translated into the first row of scaling matrix, adjusting its extreme negative alpha much closer to its true value of *GIR* = -0.09 under the true model.

The top three rows of Panel C shows $\alpha^*$, *IR* and *GIR* for each of the five portfolios under both *FF3* and *Carhart4*. $\alpha^*$ is normalized by the cross-sectional average standard deviation (i.e., the average of the square root of the diagonal elements in $\Sigma$ and $\Phi$) to bring normalized alphas to the same scale as *IR* and *GIR*, so that the three performance metrics are comparable in scale. *IR* is calculated by taking the ratio of each asset's alpha to its standard deviation (i.e., the square root of the corresponding diagonal element in $\Sigma$ and $\Phi$). We find a monotonically increasing pattern for $\alpha^*$ and *IR* under *FF3*, with *IR* having a lower dispersion than $\alpha^*$. By incorporating the heteroskedastic risk information cross sectionally, *IR* produces a more stable performance measure than the normalized alpha $\alpha^*$. More importantly, we observe that the systematic pattern of $\alpha^*$ and *IR*, which is a signal of model misspecification, disappears for *GIR*. By incorporating information in both variances and covariances in Σ, *GIR* under *FF3* ($GIR_3$) produces the same cross-sectional pattern as that under *Carhart4* ($GIR_4$). To elaborate, if the five portfolios are ranked according to their *GIRs*, one can verify that the ranking based on $GIR_3$ is identical to the ranking based on $GIR_4$. This means that, despite the use of a misspecified model for performance evaluation, the *GIR* measure produces identical performance ranking *as if* the true model were used.

To formally measure the effect of model misspecification on performance evaluation, the distance metric as in eq. (17) is presented at the bottom three rows of Panel C. The left panel shows the differences of the three performance measures between *FF3* and *Carhart4* for each portfolio, and the right panel shows the distance metric calculated as the square root of the sum of squared differences. The distance metric measures the cross-sectional dispersion of the respective measures relative to true values. If normalized alpha is the performance measure, the shortest distance between the misspecified model *FF3* and the true model *Carhart4* is 0.45. The distance is lowered to 0.35 if performance is measured by *IR*, and reduced sharply to 0.13 if measured by *GIR*, consistent with *Hypothesis 1.a*. This suggests that when a misspecified model is used to evaluate the performance of active portfolios, *GIR* is the most reliable performance measure that is much closer to the true value than the other two competing measures.

In sum, the illustrative example shows how *GIR* exploits the residual covariance information in measuring skills under a misspecified model, and the stylized results support *GIR* as a robust performance measure. Furthermore, when the true model *Carhart4* is used as the benchmarks, i.e., residual information is no longer useful, one can verify that $\alpha^*$ produces the lowest cross-sectional dispersion (by taking the square root of summed squares for the first three rows of Panel C under *Carhart4*). That is, *GIR* and *IR* lose their information content under the true model, so *Hypothesis 1.c* is observed.

## 3. Simulations

To compare the relative performance of $\alpha^*$, *IR*, and *GIR*, we employ a simulation method for which the true model is known. The three performance measures under varying degrees of model misspecification are compared against those under the true model. We adopt the simulation design of Kothari and Warner (2001) who use randomly stratified portfolios to simulate fund styles but use real data to simulate fund returns. This ensures that the residual covariance matrix is determined by real unexplained returns but not by any artificially imposed structure.[12]

### 3.1 Simulation Design

Fama-French characteristics-sorted portfolios serve as the base assets for simulation for the following reasons. First, these portfolios represent a wide range of style-based actively managed funds (Sharpe, 1992, 1994), including size-, growth-, value-, momentum-, profitability-, and investment-based funds that are widely practiced in active asset management. Although these portfolios are mechanically sorted by firm characteristics, they *appear* active to the econometrician who uses the wrong benchmarks to measure their performance. In our simulation, the activeness of these portfolios can be thought of managers adopting benchmark gaming strategies (Goetzmann, Ingersoll, Spiegel, and Welch, 2007; Sensoy, 2009). For example, if a manager uses net share issues to pick stocks into the fund self-designated as a growth or value style, using *FF3* or *Carhart4* as the benchmark model is likely to find large magnitude of alphas.[13] As such, the wide range of Fama-French portfolios creates large spreads of abnormal returns that can represent the skills of outperforming and underperforming managers. Second, these portfolios are constructed to capture the failure of well-known asset-pricing models (factors) including the *CAPM* (*MKT*), the Fama-French three-factor model *FF3* (*MKT SMB HML*), Carhart's (1997) four-factor model *Carhart4* (*MKT SMB HML UMD*). In addition, the Fama-French five-factor model *FF5* (*MKT SMB HML RMW CMA*) can explain a wide array of anomalies. However, *FF5* fails to explain momentum (Fama and French, 2016; Hou, Xue, and Zhang, 2016). Thus, the four benchmark models (*CAPM*, *FF3*, *Carhart4*, and *FF5*) are known to be misspecified to a diminishing degree to evaluate the performance of style-based portfolios. Third, the six-factor model, i.e., *FF5* plus the momentum factor dubbed *FF6* (*MKT SMB HML RMW CMA UMD*), is known to capture cross-sectional average returns of the wide range of portfolios (Fama and French, 2016, 2017); hence *FF6* serves as the true model for performance evaluation of style-based investment funds. Even under the true model *FF6*, Fama and French (2015, 2016) show that a number of portfolios still cannot be explained; for example, the portfolio of unprofitable small firms that invest a lot exhibits a significantly negative alpha. These portfolios mimic truly outperformed or underperformed funds in our simulation.

---

[12] An alternative simulation design used by MacKinlay and Pastor (2000) and Cohen, Covar, and Pastor (2005) fully specifies the underlying return-generating process of fund returns. However, this design needs to impose artificial structures on the residual covariance matrix. We also conduct similar simulation procedures by line spacing the factor exposures to generate the systematic part of fund returns; for the residual returns, we impose a correlation structure for randomly selected subgroups to be correlated to varying degrees. The results (untabulated) essentially support the main conclusion tabulated using the Fama-French real portfolios which give more direct evidence.

[13] Using benchmark gaming to motivate active strategies is admittedly stylized, e.g., no fund can game the benchmark for as long as 30 years of evaluation period. However, the purpose of the simulation is not to mimic real active strategies, but rather to examine the effects of omitted factors on fund performance. This design also ensures that the three performance measures are compared on the same level playing field.

The universe of style-based investment funds covers a wide range of assets, including four sets of bivariate-sorted portfolios (25 *Size-B/M* portfolios, 25 *Size-OP* portfolios, 25 *Size-INV* portfolios, and 25 *Size-MOM* portfolios), three sets of three-way-sorted portfolios (32 *Size-B/M-INV* portfolios, 32 *Size-B/M-INV* portfolios, and 32 *Size-OP-INV* portfolios). In addition, following Fama and French (2016), also added to the fund universe are 15 sets of univariate-sorted decile portfolios that cover a board array of anomalies, most of which are not targeted by the passive benchmarks. These 15 decile portfolios are classified into four groups: 1) FF-factor group containing 40 decile portfolios formed on market capitalization (*Size*), book-to-market (*B/M*), profitability (*OP*), and investment (*INV*); 2) valuation group containing 30 decile portfolios formed on earnings-to-price (*E/P*), cash flow-to-price (*CF/P*), and dividend yield (*D/P*); 3) prior return group containing 30 decile portfolios formed on momentum (*MOM*), short-term reversal (*STR*), and long-term reversal (*LTR*); and 4) other anomaly group containing 50 decile portfolios formed on accruals (*AC*), net share issues (*NI*), market beta (*beta*), variance (*VAR*), and residual variance (*RVAR*). In sum, the fund universe contains 4*25 double-sorted portfolios, 3*32 triple-sorted portfolios, plus 150 decile portfolios, totaling 346 portfolios that represent a variety of style-oriented investment strategies. We seek to explain excess returns on randomly selected subsets of the 346 active funds.

Our simulation procedure extends the design of Kothari and Warner (2001) to a pool of multiple portfolios at a given month. Specifically, for each month $t$, we randomly select without replacement $n$ = 10, 25, 50, 100, 200 number of funds from the universe of 346 Fama-French portfolios. To the econometrician who does not know the true factors that generate returns, each portfolio represents a style-based active fund and has an equal likelihood to be selected. Once selected, the pool of funds is held for $L$ = 36, 60, 120, 240, 360 months for performance evaluation at the end of the period. The length of evaluation period mimics varying sample size of available fund histories. Denote the intersect of the number of funds and the length of evaluation period by $n$ and $L$.[14] For example, $n = 25$ and $L = 120$ indicate

[14] $n$ is required to be strictly smaller than $L$ so that the residual covariance matrix is invertible. Even when $n$ is smaller but close to $L$, the estimate of the residual covariance matrix has a significant amount of sampling error, and its inverse is a poor

that for each month $t$ starting 1963:07, 25 funds are randomly selected from the universe of 346 Fama-French portfolios, and are held for 120 months till 1973:06 when the pool of funds are evaluated and ranked. For the next month 1963:08, a new pool of 25 funds is randomly created and held for 120 months till 1973:07 for evaluation. This rolling window is repeated each month until $t$ = 2007:01, and the 25 funds are held for 120 months till 2016:12, the end of the sample period. Thus, this simulation procedure creates 522 overlapping samples of randomly stratified sets of active funds whose performance is evaluated and ranked at the end of each evaluation period.

The data on passive benchmarks include monthly returns on the six factors (*MKT SMB HML RMW CMA UMD*). The full sample is from 1963:07 to 2016:12, totaling 642 monthly observations. We consider subsamples of varying lengths to simulate the available histories of fund returns. All the data on base assets and passive benchmarks are from the data library of Professor French. For each fund $i$ in the cross section, its performance is evaluated by various factors as shown in the parentheses of the five models (four misspecified plus the true model) above described. For example, if the true model *FF6* is used to measure skills, the following time-series regression is performed:

$$R_{it} - R_{ft} = \alpha_i + b_i MKT_t + s_i SMB_t + h_i HML_t + r_i RMW_t + c_i CMA_t + u_i UMD_t + \varepsilon_{it}, \quad (18)$$

where $R_{it}$ is month $t$'s return on portfolio $i$ within a given set of randomly selected funds, and $t$ indicates each month of the evaluation period. $a$ and $\Phi$ are produced from time-regressions using the true model, and $\alpha$ and $\Sigma$ are produced from time-series regressions of *CAPM*, *FF3*, *Carhart4*, and *FF5*, respectively.

The vector of intercepts ($\alpha$ and $a$) and the residual covariance matrix ($\Sigma$ and $\Phi$) are obtained from the respective models throughout the overlapping samples on rolling window basis. The three performance measures ($\alpha^*$, *IR*, *GIR*) are computed for each sample. Overlapping time series of the three measures are collected and compared for various statistics and distance metrics described in Section 1.[15]

### 3.2 Monthly Performance

estimator for the inverse of the true residual covariance matrix (Bai and Shi, 2011). We first present the simulation results for these cases then propose a solution in Section 3.4.

[15] The simulation results reported in Tables 2-6 can be exactly reproduced with a random seed of 10. We also conduct multiple runs, and take the average results across a large number of runs. There is little variation across different runs, showing remarkable stability of simulation results. Therefore, results from the random seed of 10 are reported hereafter for ease of replication.

We examine the three performance measures ($\alpha^*$, *IR*, *GIR*) under differential degrees of model misspecification (*CAPM*, *FF3*, *Carhart4*, *FF5*) along two dimensions: 1) across various number of funds in the pool $n$ = 10, 25, 50, 100, and 200; and 2) across varying lengths of evaluation period $L$ = 36, 60, 120, 240, and 360 months. Below we present two views of the simulation results. In the first view, we fix $n$ to examine the performance measures across $L$. In the second view, we fix $L$ to examine the performance measures across $n$. Comparative results under different model specifications are presented under both views.

### 3.2.1 Baseline Results ($n$ = 25, $L$ = 120)

[Insert Table 2 here]

We use $n$ = 25 funds and $L$ = 120 months as the baseline setting to interpret the simulation results in details. The baseline results serve as the comparative reference for the interpretations of other results as $n$ and $L$ are varied. Panel A.3 of Table 2 (shown in bold face) reports the baseline results under *CAPM*, *FF3*, *Carhart4*, and *FF5*, which are designed to be misspecified to a diminishing degree relative to the true model *FF6*. The values in the first four columns are the time-series means of 522 overlapping observations for the distance metrics defined in eq. (17), which measure the average cross-sectional dispersion of each performance measure between the misspecified model and the true model (i.e., square root of sum squared difference). The unit of these distance values is abnormal returns per unit of residual risk and can be interpreted as the minimum cost of moving the mass of $P^*$ in eq. (11) to $Q^*$ in eq. (14). For example, $\alpha^*_{CAPM}$ = 0.76 means that if normalized alpha is used as the performance measure under *CAPM*, the average minimum cost of moving the mass of $P^* \sim N(\alpha^*, I_{25})$ to $Q^* \sim N(a^*, I_{25})$ for the set of 25 style-based investment funds is 0.76 per unit of homoscedastic residual risk. $IR_{CAPM}$ = 0.69 shows that *IR* is a better performance measure that lowers the transport cost by 0.07. Moreover, *GIR* is the best performance measure that entails the most significant cost reduction. Compared to $IR_{CAPM}$ = 0.69, $GIR_{CAPM}$ = 0.49 further cuts the cost by 0.20. The statistical significance of the three performance measures is conducted by the mean-difference $t$-test with the corresponding $t$-statistics reported in the last four columns. Under the *CAPM*, for example, $t(\alpha^* - IR)$ = 5.56, $t(IR - GIR)$ = 19.72, and $t(\alpha^* - GIR)$ = 23.68 show that the respective mean differences are all highly significant, and using *GIR* to measure performance yields the most significant improvement.

Examining the performance measures across models from *CAPM* to *FF5*, we find a monotonically decreasing pattern for all three measures. For example, the transport cost is reduced by 0.30 from $GIR_{CAPM}$

= 0.49 to $GIR_{FF5}$ = 0.19. In our controlled experiment, this is simply the evidence of diminishing degree of model misspecification from *CAPM* to *FF5*; intuitively, *FF5* is *closer* to the true model than the other three models, as verified by its smallest distance to the true model. This finding suggests that a better-specified model is indeed useful to obtain a more precise performance evaluation. More importantly, there is a consistent pattern for three performance measures across all models: *GIR* is the best measure, *IR* ranks the second, and $\alpha^*$ ranks the last as measured by the distance metrics, supporting *Hypothesis 1.a*. As explained earlier, this is because *GIR* makes full use of the residual covariance information, *IR* uses partial information, and $\alpha^*$ uses none. Furthermore, we find strong evidence consistent with *Hypothesis 2*: the more misspecified the model is, the more the benefit using *GIR* as the performance measure. This is observed by comparing the significance of the mean difference *t*-statistics for *CAPM* (the most misspecified model) with that of *FF5* (the least misspecified model). Under *CAPM*, the *t*-statistic for the mean difference between the *IR* series and the *GIR* series, $t(IR - GIR)$ is 19.72, and the mean difference between the $\alpha^*$ series and the *GIR* series, $t(\alpha^* - GIR)$ is 23.68. Both values are considerably higher than $t(IR - GIR)$ = 7.20 and $t(\alpha^* - GIR)$ = 11.77 under *FF5*, suggesting greater gains of using *GIR* to measure performance when the model is deeply misspecified.

It is also economically meaningful to cross-compare performance measures under different models, using a distance equivalence measure to explain the substitution effect. This notion is inspired by Grinblatt and Titman (1994) who compare the choice of benchmarks with the choice of models on performance evaluation, and they find that the choice of benchmarks is far more important. To illustrate the substitution effect, we show in Panel A.3, $GIR_{CAPM}$ = 0.49 is about distance equivalent to $IR_{FF3}$ = 0.50, indicating that using *GIR* under *CAPM* produces the same distance (to the true model) as using *IR* under *FF3*. This implies that the choice of performance measures can substitute the effect of the choice of models, i.e., using a robust performance measure under a poorly specified model produces the same cross-sectional dispersion (relative to the true model) as using a poor performance measure under a better specified model. There are a few other instances of distance equivalent results in Panel A.3. $GIR_{FF3}$ = 0.40 is distance equivalent to $\alpha^*_{FF4}$ = 0.41, and $GIR_{FF4}$ = 0.31 is distance equivalent to $\alpha^*_{FF5}$ = 0.31. These results further demonstrate the substitution effect of a robust performance measure for a poorly specified model.

### 3.2.2 The Effects of Varying Length of Evaluation Period

Still for $n$ = 25, with $L$ = 120 (Panel A.3) as the comparison benchmark, Panels A.1, A.2, A.4, and A.5 show the results for $L$ = 36, 60, 240, and 360 months, respectively. Examining across panels, we find that all the distance values decrease monotonically from Panel A.1 ($L$ = 36) to Panel A.5 ($L$ = 360) but at a diminishing rate. Take *FF5* for example, $GIR_{L=36}$ = 1.07, $GIR_{L=60}$ = 0.29, $GIR_{L=120}$ = 0.19, $GIR_{L=240}$ = 0.18, and $GIR_{L=360}$ = 0.17. This pattern is the result of increased precision in estimating the residual covariance matrix as the sample size of fund returns grows. In particular, when the number of funds $n$ = 25 is close to the sample size $L$ = 36, estimating the residual covariance matrix becomes highly imprecise, blowing up the *GIR* values for all the models, as indicated by the negative $t$-statistics in Panel A.1 (the solution is deferred to Section 3.4). For this case ($n$ = 25 $L$ = 36), we also find that *IR* produces lower distance values than $\alpha^*$ for all the models. These results are consistent with *Hypothesis 1.b* that information in residual covariances is dominated by estimation error, but residual variances still contain valuable information in measuring skills. However, the increase in sample size effectively addresses the estimation problem. With $n$ fixed at 25, the larger the sample size $L$, the more significant the benefit of using the *GIR* measure. Such benefits can be intuitively evaluated by the distance equivalent measure of the substitution effect. For example, for $n$ = 25 and $L$ = 360 in Panel A.5, $GIR_{CAPM}$ = 0.39 is close to be distance equivalent to $\alpha^*_{FF4}$ = 0.36, and $GIR_{FF3}$ = 0.32 is distance equivalent to $\alpha^*_{FF5}$ = 0.31, suggesting a stronger substitution effect than smaller sample sizes in Panels A.1 to A.4. The stronger substitution effect is also confirmed by the more significant mean difference $t$-statistics, as compared to those in other panels.

With reference to the above baseline results for $n$ = 25, below we discuss the results of $n$ = 10, 50, 100, and 200 in Panel B, C, D and E of Table 2, respectively, as the length of evaluation period $L$ increases from 36 to 360 months in sub-panels. For all the panels, we observe patterns consistent with the baseline results: there are generally significant gains of using *GIR* compared to using $\alpha^*$ and *IR* to measure performance, confirming *Hypothesis 1.a*. Below we first examine the effect of increased precision of estimating residual covariance matrix as the sample size $L$ enlarges. We find that, in each panel, the distance values monotonically decrease for all three performance measures across all four models as the length of evaluation period increases; and the more misspecified the model is, the more significant the distance values decrease. For example, in Panel B ($n$ = 10), $GIR_{CAPM}$ = 0.71 when $L$ = 36 months reduces to $GIR_{CAPM}$ = 0.31 when $L$ = 360 months, a decrease of 0.40; in comparison, $GIR_{FF5}$ = 0.25 when $L$ = 36 months reduces to $GIR_{FF5}$ = 0.12 when $L$ = 360 months, a decrease of 0.13. These results are consistent with *Hypothesis 2*. We next examine the effect of the number of funds on performance measure. The general finding is that the gains of using *GIR* become greater as the number of funds gets larger. We can use the distance equivalence measure to evaluate the substitution effect. The most significant substitution effect is shown in Panel E.2 for $n$ = 200 and $L$ = 360, i.e., $GIR_{CAPM}$ = 0.49 is smaller than $IR_{FF5}$ = 0.67 and $\alpha^*_{FF5}$ = 0.94. This suggests that using *GIR* under the most misspecified model (*CAPM*) performs even better than using the other two measures under the least misspecified model (*FF5*). In comparison, when the number of funds is small, i.e., $n$ = 10 and $L$ = 360 in Panel B.5, the benefit of choosing *GIR* over $\alpha^*$ and *IR* is still highly significant, as indicated by the mean-difference $t$-statistics; however, the benefit is not as large as when $n$ is large. Finally, in the cases where the number of funds $n$ is close to the length of evaluation period $L$, i.e., $n$ = 25 $L$ = 36, $n$ = 50 $L$ = 60, $n$ = 100 $L$ = 120, we find that the distance metrics produced by *GIR* blow up to meaningless values. This is due to an estimation problem that the amount of noise in the estimated residual covariances outweighs the information content. We further examine this issue and propose a simple solution in Section 3.4. However, in all these cases, residual variances still contain useful information, making *IR* a shorter distance than $\alpha^*$, i.e., $t(\alpha^* - IR)$ is always significantly positive. Results in these cases support *Hypothesis 1.b.*

Figure 1 plots time-series patterns of the distance metrics produced by $\alpha^*$, *IR* and *GIR*, with each model as a separate plot, for the case of $n$ = 200 and $L$ = 360. The four plots allow us to visualize, for each month, the maximum distance discrepancy between the three performance measures produced by our simulation. In each plot, we find that $\alpha^*$ and *IR* series vary with each other closely, and both series exhibit a considerable magnitude of instability over time. In contrast, the *GIR* series lies far below the $\alpha^*$ and *IR* series, and is much stable over time. The time-series plots, coupled with the summary statistics in Panel E of Table 2, provide strong evidence for the maximum potential gains for using *GIR* as a robust performance measure.

[Insert Figure 1 here]

### 3.2.3 The Effects of Varying Number of Funds

Above subsection focuses on the effect of varying $L$ on the performance measures for each $n$. This subsection changes the view by fixing $L$ and examining the effects of varying $n$ on the performance measures.

To do this, we first propose a slightly modified distance metric. Note that the distance metric in Tables 1 & 2 involves transporting the distribution of the entire cross section (namely, Total Distance or *TD*), so the distance values become larger as the number of funds increases. For example, for $L$ = 360 in each panel of Table 2, $GIR_{FF5}$ = 0.12 for $n$ = 10, $GIR_{FF5}$ = 0.17 for $n$ = 25, $GIR_{FF5}$ = 0.20 for $n$ = 50, $GIR_{FF5}$ = 0.22 for $n$ = 100 and 200. We notice a decreasing rate of increase in distance values as $n$ enlarges. To evaluate performance measures across different number of funds, we define an *Average Distance* (*AD*) as the square root of *mean* squared difference, i.e., by dividing the distance metrics in eq. (17) by the square root of $n$. This *AD* metric standardizes the number of funds so that fund pools with different $n$ can be compared by the average distance per fund. Also note that the mean-difference $t$-statistics for *AD* do not change from those for *TD* in Table 2.

[Insert Table 3 here]

Table 3 shows the values of average distance across different number of funds by fixing $L$ = 36 in Panel A, $L$ = 60 in Panel B, $L$ = 120 in Panel C, $L$ = 240 in Panel D, and $L$ = 360 in Panel E. We find that the average distance values of $\alpha^*$ and *IR* generally remain constant or exhibit a slight increase pattern as $n$ grows. Take $L$ = 120 in Panel C for example, $\alpha^*_{CAPM}$ = 0.15 for $n$ = 10 and 25, 0.16 for $n$ = 50 and 100; $IR_{CAPM}$ = 0.14 for $n$ = 10, 25, 50 and 100. $\alpha^*$ and *IR* for other models show a similar pattern. In contrast, *GIR* clearly exhibits a decreasing pattern as $n$ increases except for a few cases where $n$ is close to $L$ (see Section 3.4 for the explanation). This is best exemplified for $L$ = 360 in Panel E, where $GIR_{CAPM}$ = 0.10 for $n$ = 10, 0.08 for $n$ = 25, 0.06 for $n$ = 50, 0.05 for $n$ = 100, and 0.03 for $n$ = 200. *GIR* for other models and in other panels generally exhibits similar pattern, with the exception for $L$ = 240 in Panel D where the distance values of $n$ = 200 appear larger than those of $n$ = 100 (see Section 3.4 for the explanation). These results confirm findings in the last subsection that using *GIR* as the performance measure is more beneficial when the number of funds is large. Such benefits can be evaluated by the distance equivalence measure for the substitution effect. To illustrate, note that, in each panel, the average distance decreases sharply across models from *CAPM* to *FF5*, i.e., the result of diminishing degree of model misspecification. We now fix the length of evaluation period at $L$ = 360 shown in Panel E. When the number of funds is small at $n$ = 10, $GIR_{CAPM}$ = 0.10 is distance equivalent to $\alpha^*_{FF3}$ = 0.11. When $n$ = 50, $GIR_{CAPM}$ = 0.06 is distance equivalent to $\alpha^*_{FF5}$ = 0.06, suggesting that using *GIR* as the performance measure can substitute the effect of a poorly misspecified model. In particular, when the number of funds is large, *GIR* under the most misspecified model (*CAPM*) produces an even closer distance to the true model than using *IR* under the least misspecified model (*FF5*), i.e., $GIR_{CAPM}$ = 0.05 and 0.03 versus $IR_{FF5}$ = 0.05 and 0.05 for $n$ = 100 and 200. These results simply present a different view from Table 2 to support *Hypotheses 1.a* and *2*.

### 3.3 Out-of-Sample Performance

In this section, we explore the persistence in alphas for simulated funds ranked on the three performance measures. The simulation procedure is similar to that described in the previous section. Specifically, for each month $t$, we randomly select without replacement $n$ = 10, 25, 50, 100, 200 number of funds from the universe of 346 Fama-French portfolios. Then the set of selected funds is held for $L$ = 36, 60, 120, 240, 360 months. At the end of the evaluation period, we rank the funds by $\alpha^*$, *IR*, and *GIR*, respectively, and pick the top quintile and bottom quintile funds based on each performance measure. Then, a long-short portfolio is formed by taking long positions on top quintile funds and short positions on bottom quintile funds, and equal-weighted portfolio returns are computed over the following (out-of-sample) year, and alphas are computed from time-series regressions for this year. Next, we roll over one month and repeat the procedure until the end of the sample. Finally, we compute time-series average alpha $A(\alpha)$, time-series standard deviation $\sigma(\alpha)$, and time-series $t$-statistic $t(\alpha)$ of these alphas over all the overlapping out-of-sample years. The standard errors in the $t$-statistics are adjusted for serial correlations using the Newey and West (1987) method with the time lag of 5 used by Kothari and Warner (2001).

[Insert Table 4 here]

Table 4 reports the out-of-sample performance results based on $\alpha^*$, *IR*, and *GIR* rankings under *CAPM*, *FF3*, *Carhart4*, and *FF5* across $n$ = 10, 25, 50, 100, 200 number of funds and $L$ = 36, 60, 120, 240, 360 months of evaluation period. The values of $A(\alpha)$, $\sigma(\alpha)$, and $t(\alpha)$ are shown on three lines of each panel. We first illustrate the time-series stability of alphas from the values of $\sigma(\alpha)$ on the second line across all panels. The most noticeable finding in Table 4 is that the long-short portfolio ranked by *GIR* consistently produces the lowest $\sigma(\alpha)$ across different models, different length of evaluation period, and different number of funds. This demonstrates a remarkable stability of using *GIR* to identify skilled funds. Further, the discrepancy of alpha standard deviation between $\alpha^*$, *IR*, and *GIR* is positively associated with the number of funds. The larger the number of funds, the more stable the out-of-sample alphas ranked by *GIR*. For example, examining $L$ = 240 under *FF5* for $n$ = 10, we find that $\sigma(\alpha)$ produced by the $\alpha^*$, *IR*, and *GIR* rankings is 0.93%, 0.87%, and 0.84%, respectively, i.e., only a small discrepancy between each other. As $n$ increases to 25, 50, 100, and 200, $\sigma(\alpha)$ under all three rankings monotonically decreases because the long-short portfolios become increasingly diversified with more funds in the top and bottom quintiles. More importantly, it is $\sigma(\alpha)$ ranked by *GIR* that decreases in much greater magnitude than $\sigma(\alpha)$ ranked by $\alpha^*$ and *IR* as $n$ enlarges. For $n$ = 200, $\sigma(\alpha)$ produced by the $\alpha^*$, *IR*, and *GIR* rankings is 0.38%, 0.34%, and 0.15%, respectively, with the last value less than half of the first two. These results suggest that fund performance becomes more stable when a large number of funds are pooled together for evaluation. This is because the large pool makes the most efficient use of the residual covariance information, and suggests that the exploiting the $\alpha$-$\Sigma$ link can help identify funds that perform much less volatile out of sample.

We next examine the average alpha $A(\alpha)$ on the first line across panels of Table 4, with the focus on benchmark stability of performance measures. The benchmark stability of a performance measure refers to its robustness to a change in the passive benchmarks. Such a notion of stability can be intuitively captured by the cross-model standard deviation of $A(\alpha)$, denoted by $std(\alpha)$, which measures the dispersion of $A(\alpha)$ across different models. We find that except for one or two minor exceptions, the *GIR* ranking produces average alphas that are the least dispersed across models. For example, in the panel of $n$ = 25 and $L$ = 120, $std(\alpha)$ is 0.137% if ranked by $\alpha^*$, 0.11% if ranked by *IR*, and 0.076% if ranked by *GIR*. The benchmark stability is more pronounced as the number of funds increases. In particular, when $n$ reaches 200, the *GIR* ranking produces $std(\alpha)$ two to three times smaller than that produced by the $\alpha^*$ and *IR* ranking. The benchmark stability presents out-of-sample evidence that *GIR* is the least susceptible to the choice of models.

Finally, in each panel of Table 4, the third line shows the out-of-sample $t$-statistic $t(\alpha)$, computed as $A(\alpha)$ divided by standard errors adjusted for serial correlations by the Newey and West method. $t(\alpha)$ measures the ability of the performance measure to identify significant fund skills, given a specific benchmark model to measure abnormal returns. In the majority of the panels, we find that using *GIR* to rank top-quintile and bottom-quintile funds produces the highest $t$-statistics. Once again, the performance of *GIR* relative to $\alpha^*$ and *IR* depends on the number of funds included in the evaluation pool. When the number of funds is large, $n$ = 100 or 200, *GIR* consistently produces the highest $t$-statistics across all models. Furthermore, we find that the highest $t$-statistics are mainly attributed to the low alpha standard deviation produced by *GIR*. As seen earlier, when $n$ = 100 or 200, $\sigma(\alpha)$ under *GIR* is typically twice or three times lower than that under $\alpha^*$ and *IR*.

To summarize our simulation results on performance measures based on monthly returns, we find that *GIR* produces the most stable measure of fund performance in terms of both in-sample distance metrics and out-of-sample portfolio returns of selected funds. The stability results suggest that *GIR* is robust to benchmark error or correlated residuals induced by any form of manager covariates. The potential gains of using *GIR* are greater when the model is more misspecified, when the number of funds under evaluation is larger, and when the length of evaluation period is longer. As for the other two measures, we find that *IR* consistently outperforms $\alpha^*$ (i.e., ratio-based normalized alphas produce the same performance ranking as return-based alphas, see Footnote 11) throughout all the simulation results.[16] That is, *Hypothesis 1.c* is not supported by any simulation results. Overall, our monthly results strongly support making full use of residual covariance information in measuring skills.

### 3.4 Weekly Performance

From previous subsections, we notice unstable results using the *GIR* measure when the length of evaluation period is short, i.e., $L$ = 36 months. In particular, when the number of funds is close to fund available histories, i.e., $n$ = 25 $L$ = 36, $n$ = 50 $L$ = 60, $n$ = 100 $L$ = 120, the distance metric blows up to meaningless values. These are the cases for which simulation results are consistent with *Hypothesis 1.b* instead of *1.a*. These unpleasant results are due to a large amount of sampling error in the estimate of the inverse of the residual covariance matrix when $n$ is close to $L$. Bai and Shi (2011) show that even though the sample estimate (denoted by $S$) is an unbiased estimator of the population Σ, its inverse $S^{-1}$ can be a poor estimator for $\Sigma^{-1}$ (i.e., the inverse needs to be performed to compute *GIR*). Under the normality assumption required to derive *GIR*, the expected value of the inverse is given by $E[S^{-1}] = \frac{L}{L-n-2} E[\Sigma^{-1}]$. In the monthly simulation, $E[S^{-1}] = 4E[\Sigma^{-1}]$ for $n$ = 25 $L$ = 36; $E[S^{-1}] = 7.5E[\Sigma^{-1}]$ for $n$ = 50 $L$ = 60; and $E[S^{-1}] = 6.7E[\Sigma^{-1}]$ for $n$ = 100 $L$ = 120. In these cases, estimation error overwhelmingly outweighs the information content embedded in the residual covariance matrix. A rule of thumb is to require $L$ to be at least twice as large as $n$ ($L \geq 2n$), because when $n = L/2 + 2$, $E[S^{-1}] = 2E[\Sigma^{-1}]$ (Bai and Shi, 2011). One can verify that when this condition satisfies, our simulation always produces expected results consistent with *Hypothesis 1.a*.

In this subsection, we aim to address the problem of estimation precision. The solution lies in finding a more accurate estimate of the residual covariance matrix. Fortunately, we learn from Merton (1980) that: 1) the variance of returns can be estimated far more precisely from historical observations than can the expected return; 2) the accuracy of the variance estimator can be improved by choosing finer observation intervals for a fixed length of the observation period. These theoretical results motivate us to increase the data frequency to weekly returns for $L$ = 36 months for which the shortest sample size creates the most severe estimation error. This is also in the spirit of Busse (1999) who shows that analyzing daily fund returns can uncover manager skills not evident in monthly returns. However, unlike his use of higher frequency data to estimate time-varying volatility, our purpose is to address the problem of estimating high dimensional covariance matrices.

[16] *IR* is closely related to alpha *t*-statistic, e.g., see Goodwin (1998). Our simulation result is consistent with the finding of Kosowski, Timmermann, Wermers, and White (2006) that ranking by the alpha *t*-statistic is more effective than by alphas in identifying skilled managers.

We select a universe of 175 Fama-French portfolios for which daily returns are provided from the French Data Library. They include 25 *Size-B/M* portfolios, 25 *Size-OP* portfolios, 25 *Size-INV* portfolios, 25 *B/M-OP* portfolios, 25 *B/M-INV* portfolios, 25 *OP-INV* portfolios, and 25 *Size-MOM* portfolios. Daily returns are cumulated to compute weekly returns for these 175 portfolios and also for the six factors (*MKT SMB HML RMW CMA UMD*) used as passive benchmarks. We conduct the same simulation procedure as for the monthly data with $L$ = 156 weeks (36 months) for $n$ = 10, 25, 50, 75, and 100 number of funds. Simulations start from the first week of 1963:07 on a weekly rolling-window basis until the evaluation sample reaches the last week of 2016:12. At each week, $n$ = 10, 25, 50, 75 and 100 number of funds are randomly selected from the investment universe, and each set of selected funds are held for 156 weeks for performance evaluation using $\alpha^*$, *IR*, and *GIR* measures under four models (*CAPM, FF3, Carhart4, FF5*) that are misspecified to a diminishing degree. This simulation procedure produces 2,632 overlapping time-series values of $\alpha^*$, *IR*, and *GIR* for each fund in the selected set, based on which the distance metric of eq. (17) is calculated between each model and the true model *FF6*. Table 5 shows the Total Distance (for all funds in the set), Average Distance (per fund in the set), and mean-difference $t$-statistic (between measures) of each measure under the four models. The weekly results uncover many stylized facts not found from monthly results for $L$ = 36 months. First, following the rule of thumb ($L \geq 2n$), the length of sample size ($L$ = 156 weeks) enables up to 78 funds to be included for evaluation, whereas the maximum number of funds should not exceed 18 for $L$ = 36 months. We therefore expect a much stronger substitution effect from the weekly results. This is evident from the increasingly stronger substitution effect for Panel A ($n$ = 10) to D ($n$ = 75). For example, using the Total Distance (*TD*) metric, when $n$ = 25 (Panel B), $GIR_{CAPM}$ = 0.33 substitutes $\alpha^*_{FF3}$ = 0.32, and $GIR_{FF3}$ = 0.23 for $IR_{Carhart4}$ = 0.24; when $n$ = 50 (Panel C), $GIR_{CAPM}$ = 0.41 substitutes $IR_{FF3}$ = 0.40; when $n$ = 75 (Panel D), $GIR_{CAPM}$ = 0.48 substitutes $\alpha^*_{Carhart4}$ = 0.47. The mean-difference $t$-statistics also support the stronger substitution effect as $n$ becomes larger. However, we also observe that the performance of *GIR* deteriorates for $n$ = 100 in Panel E as compared to $n$ = 75 in Panel D. This reflects the above described problem of estimating high dimensional covariance matrix as $n$ is close to $L$. When the rule of thumb of $L \geq 2n$ is applied, we find that the best performance of *GIR* occurs to $n$ = 75 for $L$ = 156 weeks. This adds one more piece of evidence in favor of the ad hoc $L \geq 2n$ rule. Second, examining the Average Distance (*AD*) metric across Panel A to D, we find a slight increasing trend of the *AD* values produced by $\alpha^*$ and *IR*; in contrast, the *AD* values produced by *GIR* exhibits a monotonically decreasing pattern, i.e., 0.079 ($n$ = 10), 0.066 ($n$ = 25), 0.058 ($n$ = 50), and 0.056 ($n$ = 75). These results further support that the more funds included for evaluation, the greater the gains of using *GIR* (as long as $L \geq 2n$). Finally, examining the mean-difference $t$-statistic, we find weekly evidence consistent with *Hypothesis 2*. *GIR* attains its greatest gain for *CAPM*, *FF3* the second, *Carhart4* the third, *FF5* the least, in the same order of the degree of model misspecification.

[Insert Table 5 here]

Table 6 presents the out-of-sample performance on the long-short portfolio of top and bottom quintile funds, following the same monthly procedure. The out-of-sample period for holding the long-short portfolio is 26 weeks (6 months). We find overwhelming evidence that, in all Panels A – E under all models (except for one minor case of $n$ = 10 under *FF5*), the portfolio of selected funds based on the *GIR* measure achieves the lowest alpha standard deviation $\sigma(\alpha)$ and the highest $t$-statistic $t(\alpha)$. Furthermore, *GIR* produces the lowest cross-model standard deviation $std(\alpha)$, showing the highest level of stability with respect to different choices of benchmark. All these measures of benefits become monotonically greater as the number of funds increases. It is worthwhile to emphasize that these strong results can only be obtained for very long evaluation period using monthly data, i.e., $L$ = 240 or 360 months. Therefore, this section shows that increasing the return sampling frequency can be very effective for using the *GIR* measure.

[Insert Table 6 here]

## 4 Conclusion

Active managers should be rewarded for alphas not for betas. By its own definition, alphas measure true skills only if all the correlated effects are controlled for. However, the performance evaluation literature has documented numerous sources of correlated residuals due to inefficient benchmarks (Roll, 1978), manager covariates (Wermers, 2011), and benchmark gaming (Sensoy, 2009), making alphas a biased measure of skills under misspecified models. This paper proposes $GIR = \Sigma^{-1/2}\alpha$ as a robust performance measure without identifying the sources of correlated residuals, in the spirit of "learning across funds" of Jones and Shanken (2005). The mechanism that makes *GIR* robust to model misspecification lies in its alpha adjustments by the scaling matrix $\Sigma^{-1/2}$, which de-correlates commonalities in residual returns and standardizes residual variances to an identity matrix. That is, *GIR* translates information embedded in the residual covariance matrix into alphas adjusted for residual risk.

Thus, so long as the magnitude of alphas is positively linked to the residual covariance matrix, large alphas and their cross products (left unexplained by a poor model) will be adjusted more heavily, thereby mitigating the correlated effects. Fortunately, the systematic $\alpha$-$\Sigma$ link must exist in theory by the arbitrage argument of MacKinaly (1995) and the performance-flow relation of Berk and Green (2004). Consistent with the theory-implied hypotheses, we demonstrate substantial gains of *GIR* over existing measures (*IR* and alphas) that do not account for information in residual correlations. The potential gains are greater for large number of funds under poorly specified models. In such cases, *GIR* under the *CAPM* is even closer to its true value than *IR* under the Fama-French five-factor model; and the out-of-sample volatility of *GIR* ranked portfolios is less than half or one third of those ranked by *IR* or alphas. The robustness of *GIR* implies that the choice of benchmarks, existence of manager covariates, and managers' incentive to game the benchmark, though still important, matter much less in measuring and ranking a pool of active funds. It helps alleviate model error concerns of practitioners and investors alike in active management.

The proper use of *GIR* can complement existing alphas-based studies that focus exclusively on identifying the sources of common variation in fund returns. For example, the literature has suggested a careful choice of benchmark models (Wermers, 2011), active peer benchmarks (Hunter, Kandel, Kandel, and Wermers, 2011), and non-benchmark assets (Pastor and Stambaugh, 2002), among others, to capture correlated residuals. In performance evaluation, one should first try these approaches to obtain a good model; the importance of a good model has been shown throughout our simulation results. However, since no model is perfect and managers may co-vary or game in any unidentified forms, one can then rely on the model-free *GIR* measure to account for residual correlations after a careful model choice. If the model is indeed well specified and claims to capture all the investment styles and manager covariates, which means that any residual correlations are random due to sampling error, then one should not observe any systematic pattern nor benefit of using *GIR* over *IR* or alphas to measure and rank the pool of funds.

There are ample research directions to follow. The simulation in this paper is specifically designed to omit certain factors and the simulated results are stylized so may not reflect true fund styles or real performance. To what extent *GIR* can generate robust measure and stable ranking for real mutual funds is a natural extension. How *GIR* performs for hedge funds and other institutional funds whose returns significantly deviate from normal distributions (i.e., the assumption required to derive *GIR*) remains an open question. Finally, despite its ease of computation, *GIR* has the stringent requirement for a long history of fund returns $L$ relative to the number of funds $n$ (i.e., a rule of thumb of $L \geq 2n$ is proposed ad hocly).

This paper advocates increasing the sampling frequency and demonstrates its effectiveness using weekly returns. However, this approach does not work for those funds whose higher frequency returns are unavailable. In addition, when $n$ is larger than $L$, the $n \times n$ inverse of the residual covariance matrix does not exist. In light of the measurement difficulty in these applications, this paper calls for a revisit of econometric work along the lines of Lediot and Wolf (2003) and Bai and Shi (2011) for accurate estimation of high-dimensional covariance matrix.

## References

Bai, J., and Shi, S., 2011. "Estimating High Dimensional Covariance Matrices and its Applications". *Annuals of Economics and Finance* 12 (2), 199-215.

Berk, J.B., and Green, R.C., 2004. Mutual Fund Flows and Performance in Rational Markets. *Journal of Political Economy* 112, 1269–1295.

Busse, J. 1999. Volatility Timing in Mutual Funds: Evidence from Daily Returns. *Review of Financial Studies* 12,1009–1041.

Carhart, M.M., 1997. On persistence in mutual fund performance. *Journal of Finance* 52, 57-82.

Chevalier J., and Ellison G. 1999. Are Some Mutual Fund Managers Better Than Others? Cross-sectional Patterns in Behavior and Performance. *Journal of Finance* 54, 875–99.

Cohen L, Frazzini A, and Malloy C. 2008. The Small World of Investing: Board connections and Mutual Fund Returns. *Journal of Political Economy* 116, 951–79.

Cohen R, Coval J, and Pastor L. 2005. Judging Fund Managers by the Company They Keep. *Journal of Finance* 60, 1057–96.

Daniel, K, M. Grinblatt, S. Titman, and R. Wermers, 1997, Measuring Mutual Fund Performance with Characteristics-based Benchmarks, *Journal of Finance* 52, 1035-1058.

Deadman, Edvin, Nicholas J. Higham, and Rui Ralha, 2013, "Blocked Schur Algorithms for Computing the Matrix Square Root", *Lecture Notes in Computer Science*, 7782. pp. 171-182.

Dawson, D. C., and B.V., Landau, 1982, The Frechet Distance between Multivariate Normal Distributions, *Journal of Multivariate Analysis* 12, 450-455.

Elton, E., M. Gruber, C. Blake, 2003, Incentive Fees and Mutual Funds, Journal of Finance 58 (2), 779 – 804.

Fama, Eugene F., and Kenneth R. French. 1993. "Common Risk Factors in the Returns on Stocks and Bonds", *Journal of Financial Economics* 33, 3–56.

Fama, Eugene F., and Kenneth R. French, 2015, A five-factor asset pricing model, *Journal of Financial Economics* 116, 1-22.

Fama, Eugene F., and Kenneth R. French, 2016, Dissecting anomalies with a five-factor model, *Review of Financial Studies*, Vol. 29 (1), 69-103.

Fama, Eugene F., and Kenneth R. French, 2018, Choosing Factors, *Journal of Financial Economics* forthcoming.

Galichon, Alfred, 2016, Optimal Transport Methods in Economics, Princeton University Press.

Galichon, Alfred, 2017, A Survey of Some Recent Applications of Optimal Transport Methods to Econometrics, *The Econometrics Journal*, doi:10.1111/ectj.12083

Givens, Clark R., and Shortt, Rae Michael, 1984, A class of Wasserstein metrics for probability distributions. *Michigan Math. J.* 31 (2), 231–240.

Goetzmann W, Ingersoll J, Spiegel M, and Welch I, 2007, Portfolio Performance Manipulation and Manipulation-proof Performance Measures, *Review of Financial Studies* 20,1503–1546.

Goodwin, T. H., 1998, The Information Ratio, *Financial Analysts Journal* 54 (4), 34–43.

Gottesman A, and Morey MR. 2006. Manager Education and Mutual Fund Performance. *Journal of Empirical Finance* 13:145–82.

Grinblatt, M. and Titman, S., 1989, Mutual Fund Performance: An Analysis of Quarterly Portfolio Holdings, *Journal of Business* 62, 394-415.

Grinblatt, M. and Titman, S., 1994, A Study of Monthly Mutual Fund Returns and Performance Evaluation Techniques, *Journal of Financial and Quantitative Analysis* 29, 419-444.

Grinblatt M, Titman S, and Wermers R. 1995. Momentum Investment Strategies, Portfolio Performance, and Herding: A Study of Mutual Fund Behavior. *American Economic Review*, 85 (10), 88–105.

Grinold, R. C. and Kahn, R. N., 1992, Information Analysis, *Journal of Portfolio Management* 18 (3), 14–21.

Grinold, R. C. and Kahn, R. N., 1995, Active Portfolio Management, Richard D. Irwin, Chicago, Illinois.

Hou, Kewei, Chen Xue, and Lu Zhang, 2015, Digesting Anomalies: An Investment Approach, Review of Financial Studies 28, 650-705.

Hou, Kewei, Chen Xue, and Lu Zhang, 2016, A Comparison of New Factor Models, Working paper.

Hunter, David, Eugene Kandel, Shmuel Kandel, and Russ Wermers, 2014, Mutual Fund Performance Evaluation with Active Peer Benchmarks, *Journal of Financial Economics* 112, 1-29.

Jones, C.S., and Shanken, J., 2005. Mutual Fund Performance with Learning across Funds. *Journal of Financial Economics* 78, 507–552.

Kosowski, R., A. Timmermann, R. Wermers, and H. White, 2006. Can Mutual Fund 'Stars' Really Pick Stocks? New Evidence from a Bootstrap Analysis. *Journal of Finance* 61 (6), 2551–2595.

Kothari, S. P. and Warner, J. B., 2001, Evaluating Mutual Fund Performance, *Journal of Finance* 56, 1985–2010.

Knott, M., and C.S. Smith, 1984, On the Optimal Mapping of Distributions, *Journal of Optimization Theory and Its Applications* 43, 39-49.

Ledoit, O. and Wolf, M., 2003, Improved Estimation of the Covariance Matrix of Stock Returns with an Application to Portfolio Selection, *Journal of Empirical Finance* 10, 603–621.

MacKinlay, A.C., 1995, Multifactor models do not explain deviations from the CAPM, *Journal of Financial Economics* 38, 3-28.

MacKinlay, A.C., and Lubos Pastor, 2000, Asset Pricing Models: Implications for Expected Returns and Portfolio Selection, *Review of Financial Studies* 13 (4), 883-916.

Merton, R.C., 1980. On estimating the expected return on the market. *Journal of Financial Economics* 8, 323–361.

Modigliani, F. and Pogue, G. A., 1974, An introduction to risk and return – II, *Financial Analysts Journal* 30 (3), 69–86.

Newey, W.D., and K. D. West, 1987, A Simple, Positive Semi-definite Heteroscedasticity and Autocorrelation Consistent Covariance Matrix, *Econometrica* 55, 703-708.

Olkin, I., and F. Pukelsheim, 1982, The Distance between Two Random Vectors with Given Dispersion Matrices, *Linear Algebra and its Applications* 48, 257-263.

Pastor, Lubos, and Robert F. Stambaugh, 2002, Comparing Asset Pricing Models: An Investment Perspective, *Journal of Financial Economics* 56, 335-381.

Pastor, Lubos, and Robert F. Stambaugh, 2002, Mutual Fund Performance and Seemingly Unrelated Assets, *Journal of Financial Economics* 63 (3), 315-349.

Pastor, Lubos, and Robert F. Stambaugh, 2015, Scale and Skill in Active Management, *Journal of Financial Economics* 116, 23-45.

Sensoy, Berk A., 2009, Performance Evaluation and Self-designated Benchmark Indexes in the Mutual Fund Industry, *Journal of Financial Economics* 92, 25-39.

Sharpe, William F., 1992, The Sharpe Ratio, *Journal of Portfolio Management* 21 (1), 49-58.

Sharpe, William F., 1994, Asset Allocation: Management Style and Performance Measurement, *Journal of Portfolio Management* 18 (2), 29-34.

Treynor, J.L., and F. Black, 1973, How to Use Security Analysis to Improve Portfolio Selection, *Journal of Business* 46, 66-86.

Villani, Cedric, 2003, Topics in Optimal Transportation, Lecture Notes in Mathematics, American Mathematical Society.

Villani, Cedric, 2009, Optimal Transport: Old and New, Grundlehren der Mathematischen Wissenschaften, 338, Springer, Berlin.

Wermers, Russ, 2011, Performance Measurement of Mutual Funds, Hedge Funds, and Institutional Accounts, *Annual Review of Finance and Economics* 3, 537-574.

# Appendix. The Distance between Two Multivariate Normal Distributions

Given two normally distributed random vectors $X\sim N(\alpha_p, \Sigma_p)$ and $Y\sim N(\alpha_q, \Sigma_q)$ in $R^n$, define the demeaned random vectors $\underline{X} = X - \alpha_p$ and $\underline{Y} = Y - \alpha_q$ . Define the squared quadratic Wasserstein distance $WD_2^2 \equiv E[||X-Y||^2] = ||\alpha_p - \alpha_q||^2 + E[||\underline{X}-\underline{Y}||^2]$. For ease of exposition, denote $||\Sigma_p - \Sigma_q|| \equiv E[||\underline{X}-\underline{Y}||^2]$. It follows that

$$WD_2 = \sqrt{||\alpha_p - \alpha_q||^2 + ||\Sigma_p - \Sigma_q||} \quad \text{(A.1)}$$

Under Gaussian measures, what remains to show is

$$||\Sigma_p - \Sigma_q|| = \mathrm{Tr}(\Sigma_p + \Sigma_q - 2\left(\Sigma_p^{1/2}\Sigma_q\Sigma_p^{1/2}\right)^{1/2}) \quad \text{(A.2)}$$

For the augmented random vector $(\underline{X}, \underline{Y})$ in $R^{2n}$, denote its covariance matrix by

$$\Psi = \begin{bmatrix} \Sigma_p & C \\ C^T & \Sigma_q \end{bmatrix} \quad \text{(A.3)}$$

Then $||\Sigma_p - \Sigma_q|| = \mathrm{Tr}(\Sigma_p + \Sigma_q - 2C)$, and the infimum of $||\Sigma_p - \Sigma_q||$ is to find $C = E[\underline{XY'}]$ so that $\underline{X}$ and $\underline{Y}$ are maximally correlated subject to the constraint that $\Psi$ is a positive definite covariance matrix. Thus, the optimization problem becomes

$$\underbrace{max}_{C} \, 2Tr(C) \quad \text{(A.4)}$$

s.t.

$$\Sigma_p - C\Sigma_q^{-1}C^T > 0 \quad \text{(A.5)}$$

where (A.5) is the Schur complement constraint.

The solution of (A.4) subject to (A.5) leads to (A.2). The detailed proof is given by Dowson and Landau (1982) and Givens and Shortt (1984), where $WD_2$ is also termed the Frechet distance.

(A.2) can also be derived from the optimal transport mapping (Knott and Smith, 1984; Olkin and Pukelsheim, 1982). To check that the optimal transport plan maps $N(\alpha_p, \Sigma_p)$ to $N(\alpha_q, \Sigma_q)$, for the zero-mean random vector $\underline{X}\sim N(0, \Sigma_p)$, let $\underline{Y} = T_p\underline{X}$, where the optimal mapping matrix is given by

$$T_p = \Sigma_p^{-1/2}\left(\Sigma_p^{1/2}\Sigma_q\Sigma_p^{1/2}\right)^{1/2}\Sigma_p^{-1/2} \quad \text{(A.6)}$$

Given $\underline{Y} = T_p\underline{X}$ and (A.6), we have

$$E[\underline{YY'}] = T_pE[\underline{XX'}]T_p' = \Sigma_p^{-\frac{1}{2}}\left(\Sigma_p^{\frac{1}{2}}\Sigma_q\Sigma_p^{\frac{1}{2}}\right)^{\frac{1}{2}}\Sigma_p^{-\frac{1}{2}}\Sigma_p\Sigma_p^{-\frac{1}{2}}\left(\Sigma_p^{\frac{1}{2}}\Sigma_q\Sigma_p^{\frac{1}{2}}\right)^{\frac{1}{2}}\Sigma_p^{-\frac{1}{2}} = \Sigma_p^{-\frac{1}{2}}\left(\Sigma_p^{\frac{1}{2}}\Sigma_q\Sigma_p^{\frac{1}{2}}\right)^{\frac{1}{2}}\left(\Sigma_p^{\frac{1}{2}}\Sigma_q\Sigma_p^{\frac{1}{2}}\right)^{\frac{1}{2}}\Sigma_p^{-\frac{1}{2}} = \Sigma_p^{-\frac{1}{2}}\left(\Sigma_p^{\frac{1}{2}}\Sigma_q\Sigma_p^{\frac{1}{2}}\right)\Sigma_p^{-\frac{1}{2}} = \Sigma_q \quad \text{(A.7)}$$

In the univariate case where $\Sigma_p = \sigma_p^2$ and $\Sigma_q = \sigma_q^2$ are scalers, the optimal mapping matrix simplifies to a scaler $T_p = \sigma_p^{-1}(\sigma_p\sigma_q^2\sigma_p)^{1/2}\sigma_p^{-1} = \sigma_q/\sigma_p$. Then $\sigma_q^2 = (T_p\sigma_p)^2$ is easily verified.

To check that $T_p$ is indeed optimal, we have

$$E[||\underline{X}-\underline{Y}||^2] = E[||\underline{X}||^2] + E[||\underline{Y}||^2] - 2E[\langle \underline{X}, \underline{Y} \rangle] = Tr(\Sigma_p) + Tr(\Sigma_q) - 2E[\langle \underline{X}, \mathrm{T}_p \underline{X} \rangle] = Tr(\Sigma_p) + Tr(\Sigma_q) - 2tr(\Sigma_p \mathrm{T}_p) = Tr(\Sigma_p) + Tr(\Sigma_q) - 2tr\left(\left(\Sigma_p^{\frac{1}{2}} \Sigma_q \Sigma_p^{\frac{1}{2}}\right)^{\frac{1}{2}}\right) = Tr(\Sigma_p + \Sigma_q - 2\left(\Sigma_p^{\frac{1}{2}} \Sigma_q \Sigma_p^{\frac{1}{2}}\right)^{\frac{1}{2}})$$

(A.8)

The second last equality is by the cyclic property of the trace operator.

The converse optimal transport mapping can also be derived. Let $\underline{X} = T_q \underline{Y}$, where the optimal mapping matrix is given by $T_q = \Sigma_q^{-1/2} \left(\Sigma_q^{1/2} \Sigma_p \Sigma_q^{1/2}\right)^{1/2} \Sigma_q^{-1/2}$. It is easy to verify that $T_q = T_p^{-1}$.

Figure 1. Distance Metrics for Three Performance Measures under Different Models
(Number of funds $n$ = 200, Length of evaluation period $L$ = 360 months)

This figure depicts the distance metrics over time for three performance measures ($\alpha^*$, *IR* and *GIR*) between a misspecified model and the true model. Four misspecified models (factors) are *CAPM* (*MKT*) in Figure 1.a, *FF3* (*MKT SMB HML*) in Figure 1.b, *Carhart4* (*MKT SMB HML UMD*) in Figure 1.c, and *FF5* (*MKT SMB HML RMW CMA*) in Figure 1.d. The true model is *FF6* (*MKT SMB HML RMW CMA UMD*). At each month starting from 1963:07, 200 funds are randomly selected from a universe of 346 Fama-French characteristics-sorted portfolios including 15×10 univariate-, 4×25 bivariate-, 3×32 triple-sorted portfolios. The selected funds are held for 360 months for performance evaluation by $\alpha^*$, *IR* and *GIR* at the end of the period. Simulation is conducted on rolling window basis until the last month of the evaluation period reaches 2016:12 (the end of sample), creating 282 overlapping time-series observations of distance metrics for $\alpha^*$, *IR* and *GIR* in each figure. All data are from the French Data Library.

The distance metric is defined as the Euclidean 2-norm $||\Sigma^{-\frac{1}{2}}\alpha - \Phi^{-\frac{1}{2}}a||$ (i.e., square root of sum squared difference), where $\alpha$ and $a$ are alphas, $\Sigma$ and $\Phi$ are residual covariance matrices, produced by the misspecified model and the true model, respectively. To compute the distance metric, *GIR* uses full information in $\Sigma$ and $\Phi$, and $\Sigma^{-\frac{1}{2}}$ and $\Phi^{-\frac{1}{2}}$ are symmetric positive-definite inverse square root of $\Sigma$ and $\Phi$, respectively; *IR* uses only diagonal information in $\Sigma$ and $\Phi$ (i.e., $\Sigma^{-\frac{1}{2}}$ and $\Phi^{-\frac{1}{2}}$ are heteroskedastic diagonal matrices); and for $\alpha^*$, $\Sigma^{-\frac{1}{2}}$ and $\Phi^{-\frac{1}{2}}$ are homoscedastic diagonal matrices whose elements are set equal to their respective cross-sectional average.

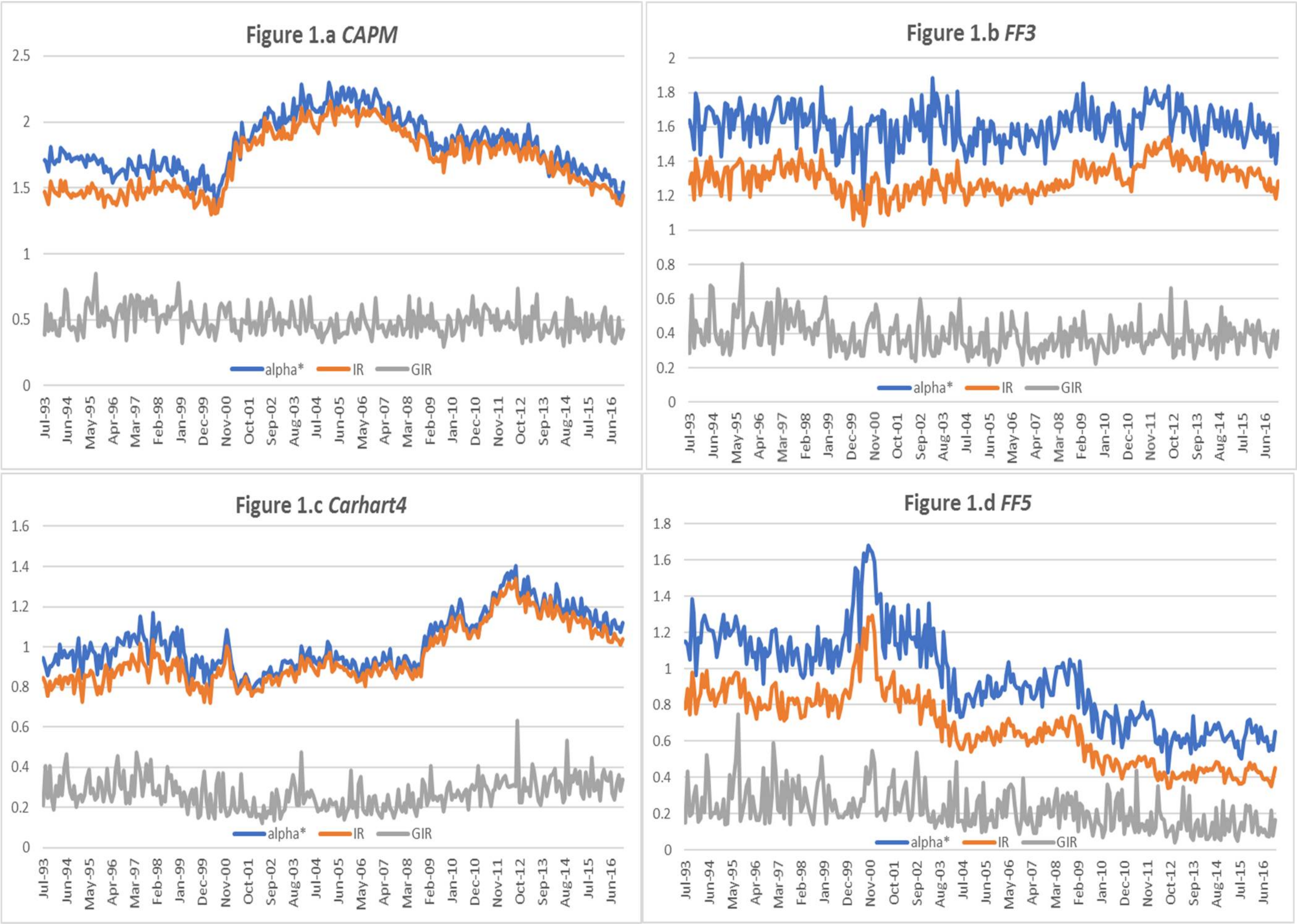

## Table 1. An Illustrative Example: Five Momentum Portfolios

This table shows the results of performance evaluation for five momentum portfolios using normalized alpha $\alpha^*$, *IR* and *GIR* measures under two benchmark models (factors): *FF3* (*MKT SMB HML*) and *Carhart4* (*MKT SMB HML UMDo*), where *UMDo* is the made orthogonal to the three factors of *FF3*. Panel A reports the coefficients (*t*-statistics) and $R^2$s from time-series regressions for the period of 1963:07 ~ 2016:12. Panel B shows the residual covariance matrix ($\Sigma$ and $\Phi$) and the optimal mapping or scaling matrix ($\Sigma^{-1/2}$ and $\Phi^{-1/2}$) under each model. Panel C shows the values of the three performance measures for each portfolio, the performance measure difference between *FF3* and *Carhart4*, and the distance metrics between the two models for each measure. Under *FF3*, the three performance measures are defined as: $GIR = \Sigma^{-1/2}\alpha$, $IR = \mathrm{D}^{-1/2}\alpha$, $\alpha^* = \alpha/\sigma_e$, where $\Sigma^{-\frac{1}{2}}$ is a symmetric positive-definite matrix as the inverse square root of $\Sigma$, $D$ uses only diagonal elements of $\Sigma$, and $\sigma_e$ is the cross-sectional average of standard errors of residuals. The distance metric is the Euclidean 2-norm $\|\cdot\|$ (i.e., square root of sum squared difference). The three measures under *Carhart4* care defined similarly with the corresponding $a$ and $\Phi$. All data are from the French Data Library.

| | ***FF3*** | | | | | ***Carhart4*** | | | | |
|---|---|---|---|---|---|---|---|---|---|---|
| | ***Panel A: Regression Results*** | | | | | | | | | |
| | *Lo PRIOR* | *PRIOR 2* | *PRIOR 3* | *PRIOR 4* | *Hi PRIOR* | *Lo PRIOR* | *PRIOR 2* | *PRIOR 3* | *PRIOR 4* | *Hi PRIOR* |
| $\alpha$ | -0.74 | -0.16 | 0.00 | 0.17 | 0.47 | -0.07 | 0.14 | 0.09 | 0.08 | 0.12 |
| (*t-stat*) | (-5.53) | (-2.44) | (0.03) | (4.27) | (6.35) | (-1.37) | (3.86) | (2.26) | (2.14) | (3.29) |
| *MKT* | 1.28 | 1.04 | 0.96 | 0.94 | 1.03 | 1.28 | 1.04 | 0.96 | 0.94 | 1.03 |
| (*t-stat*) | (40.5) | (65.8) | (94.8) | (97.2) | (58.3) | (111.6) | (122.8) | (102.9) | (107.3) | (117.2) |
| *SMB* | 0.59 | 0.44 | 0.41 | 0.42 | 0.65 | 0.59 | 0.44 | 0.41 | 0.42 | 0.65 |
| (*t-stat*) | (13.2) | (19.9) | (28.6) | (30.6) | (26.2) | (36.3) | (37.1) | (31.1) | (33.7) | (52.6) |
| *HML* | 0.24 | 0.32 | 0.29 | 0.19 | -0.17 | 0.24 | 0.32 | 0.29 | 0.19 | -0.17 |
| (*t-stat*) | (4.95) | (13.5) | (18.9) | (13.3) | (-6.47) | (13.6) | (25.1) | (20.5) | (14.6) | (-13.0) |
| *UMDo* | | | | | | -0.75 | -0.34 | -0.10 | 0.11 | 0.39 |
| (*t-stat*) | | | | | | (-64.8) | (-39.8) | (-10.7) | (11.8) | (44.1) |
| $R^2$ | 0.78 | 0.90 | 0.95 | 0.95 | 0.90 | 0.97 | 0.97 | 0.96 | 0.96 | 0.98 |
| | ***Panel B: Residual Covariance Matrix and Optimal Mapping Matrix*** | | | | | | | | | |
| | | | $\Sigma$ | | | | | $\Phi$ | | |
| Lo PRIOR | 10.86 | 4.03 | 1.00 | -1.35 | -4.27 | 1.43 | -0.22 | -0.27 | -0.03 | 0.63 |
| PRIOR 2 | 4.03 | 2.68 | 1.16 | -0.14 | -2.18 | -0.22 | 0.77 | 0.59 | 0.46 | 0.02 |
| PRIOR 3 | 1.00 | 1.16 | 1.11 | 0.43 | -0.74 | -0.27 | 0.59 | 0.94 | 0.61 | -0.08 |
| PRIOR 4 | -1.35 | -0.14 | 0.43 | 1.02 | 0.69 | -0.03 | 0.46 | 0.61 | 0.84 | 0.01 |
| Hi PRIOR | -4.27 | -2.18 | -0.74 | 0.69 | 3.37 | 0.63 | 0.02 | -0.08 | 0.01 | 0.83 |
| | | | $T_3 = \Sigma^{-1/2}$ | | | | | $T_4 = \Phi^{-1/2}$ | | |
| Lo PRIOR | 0.44 | -0.26 | 0.04 | 0.13 | 0.13 | 1.00 | 0.13 | 0.10 | -0.09 | -0.37 |
| PRIOR 2 | -0.26 | 1.26 | -0.50 | -0.07 | 0.23 | 0.13 | 1.55 | -0.46 | -0.24 | -0.14 |
| PRIOR 3 | 0.04 | -0.50 | 1.53 | -0.46 | 0.10 | 0.10 | -0.46 | 1.54 | -0.48 | 0.05 |
| PRIOR 4 | 0.13 | -0.07 | -0.46 | 1.37 | -0.20 | -0.09 | -0.24 | -0.48 | 1.47 | 0.01 |
| Hi PRIOR | 0.13 | 0.23 | 0.10 | -0.20 | 0.84 | -0.37 | -0.14 | 0.05 | 0.01 | 1.31 |
| | ***Panel C: Distance Metric for Normalized Alpha, Information Ratio, and Generalized Information Ratio*** | | | | | | | | | |
| $\alpha^*$ | -0.42 | -0.09 | 0.00 | 0.10 | 0.27 | -0.07 | 0.14 | 0.09 | 0.08 | 0.13 |
| *IR* | -0.22 | -0.10 | 0.00 | 0.17 | 0.26 | -0.06 | 0.16 | 0.09 | 0.09 | 0.14 |
| *GIR* | -0.20 | 0.09 | 0.02 | 0.06 | 0.23 | -0.09 | 0.13 | 0.04 | 0.05 | 0.17 |
| $\alpha_3^* - \alpha_4^*$ | -0.35 | -0.23 | -0.09 | 0.02 | 0.14 | $\|\|\alpha_3^* - \alpha_4^*\|\| = 0.45$ | | | | |
| $IR_3 - IR_4$ | -0.17 | -0.26 | -0.09 | 0.08 | 0.12 | $\|\|IR_3 - IR_4\|\| = 0.35$ | | | | |
| $GIR_3 - GIR_4$ | -0.11 | -0.04 | -0.02 | 0.01 | 0.05 | $\|\|GIR_3 - GIR_4\|\| = 0.13$ | | | | |

Table 2. Total Distance Metrics and Mean Difference *t* Statistics: Monthly Performance

This table shows the total distance metrics and mean difference *t*-statistics for three performance measures ($\alpha^*$, *IR* and *GIR*) between a misspecified model and the true model. Four misspecified models (factors) are *CAPM* (*MKT*), *FF3* (*MKT SMB HML*), *Carhart4* (*MKT SMB HML UMD*), and *FF5* (*MKT SMB HML RMW CMA*). The true model is *FF6* (*MKT SMB HML RMW CMA UMD*). At each month starting from 1963:07, $n$ = 25, 10, 50, 100, 200 number of funds (in Panel A, B, C, D, and E respectively) are randomly selected from a universe of 346 Fama-French characteristics-sorted portfolios including 15×10 univariate-, 4×25 bivariate-, 3×32 triple-sorted portfolios. The selected funds are held for $L$ = 36, 60, 120, 240, and 360 months (in each sub-panel) for performance evaluation by $\alpha^*$, *IR* and *GIR* at the end of the period. Simulation is conducted on rolling window basis until the last month of the evaluation period reaches 2016:12 (the end of sample), creating overlapping time-series observations of distance metrics for $\alpha^*$, *IR* and *GIR* under each model. The values reported in the first four columns are the time-series average of the distance values under each model. The values reported in the last four columns show the mean difference *t*-statistic for the distance produced by one measure relative to the distance produced by another measure. All data are from the French Data Library.

Total distance metric is defined as the Euclidean 2-norm $||\Sigma^{-\frac{1}{2}}\alpha - \Phi^{-\frac{1}{2}}a||$ (i.e., square root of sum squared difference), where $\alpha$ and $a$ are alphas, $\Sigma$ and $\Phi$ are residual covariance matrices, produced by the misspecified model and the true model, respectively. To compute the distance metric, *GIR* uses full information in $\Sigma$ and $\Phi$, and $\Sigma^{-\frac{1}{2}}$ and $\Phi^{-\frac{1}{2}}$ are symmetric positive-definite inverse square root of $\Sigma$ and $\Phi$, respectively; *IR* uses only diagonal information in $\Sigma$ and $\Phi$ (i.e., $\Sigma^{-\frac{1}{2}}$ and $\Phi^{-\frac{1}{2}}$ are heteroskedastic diagonal matrices); and for $\alpha^*$, $\Sigma^{-\frac{1}{2}}$ and $\Phi^{-\frac{1}{2}}$ are homoscedastic diagonal matrices whose elements are set equal to the inverse of cross-sectional average standard error of residuals, respectively.

| | | | | | | | | | |
|---|---|---|---|---|---|---|---|---|---|
| ***Panel A: n = 25 funds*** | | | | | | | | | |
| | ***CAPM*** | ***FF3*** | ***Carhart4*** | ***FF5*** | | ***CAPM*** | ***FF3*** | ***Carhart4*** | ***FF5*** |
| ***A.1: L = 36 months*** | | | | | | | | | |
| $\alpha^*$ | 1.35 | 0.85 | 0.68 | 0.43 | $t(\alpha^*\text{-}IR)$ | 5.35 | 3.17 | 1.85 | 3.31 |
| *IR* | 1.21 | 0.79 | 0.65 | 0.38 | $t(IR\text{-}GIR)$ | -20.11 | -20.81 | -18.82 | -16.50 |
| *GIR* | 2.49 | 2.03 | 1.73 | 1.07 | $t(\alpha^*\text{-}GIR)$ | -17.76 | -19.76 | -18.16 | -15.00 |
| ***A.2: L = 60 months*** | | | | | | | | | |
| $\alpha^*$ | 1.00 | 0.67 | 0.51 | 0.36 | $t(\alpha^*\text{-}IR)$ | 5.89 | 5.59 | 3.38 | 4.91 |
| *IR* | 0.89 | 0.60 | 0.47 | 0.30 | $t(IR\text{-}GIR)$ | 8.99 | 1.07 | -0.14 | 1.51 |
| *GIR* | 0.75 | 0.59 | 0.48 | 0.29 | $t(\alpha^*\text{-}GIR)$ | 14.02 | 6.16 | 2.98 | 6.02 |
| ***A.3: L = 120 months*** | | | | | | | | | |
| $\alpha^*$ | **0.76** | **0.59** | **0.41** | **0.31** | $t(\alpha^*\text{-}IR)$ | **5.56** | **8.95** | **4.54** | **5.72** |
| *IR* | **0.69** | **0.50** | **0.38** | **0.25** | $t(IR\text{-}GIR)$ | **19.72** | **13.45** | **11.14** | **7.20** |
| *GIR* | **0.49** | **0.40** | **0.31** | **0.19** | $t(\alpha^*\text{-}GIR)$ | **23.68** | **20.52** | **14.86** | **11.77** |
| ***A.4: L = 240 months*** | | | | | | | | | |
| $\alpha^*$ | 0.68 | 0.58 | 0.39 | 0.32 | $t(\alpha^*\text{-}IR)$ | 6.31 | 11.19 | 4.59 | 7.99 |
| *IR* | 0.63 | 0.49 | 0.36 | 0.24 | $t(IR\text{-}GIR)$ | 28.28 | 22.43 | 22.42 | 9.09 |
| *GIR* | 0.43 | 0.36 | 0.26 | 0.18 | $t(\alpha^*\text{-}GIR)$ | 32.90 | 28.39 | 26.80 | 15.57 |
| ***A.5: L = 360 months*** | | | | | | | | | |
| $\alpha^*$ | 0.64 | 0.55 | 0.36 | 0.31 | $t(\alpha^*\text{-}IR)$ | 4.19 | 9.30 | 3.11 | 6.93 |
| *IR* | 0.61 | 0.46 | 0.34 | 0.23 | $t(IR\text{-}GIR)$ | 29.76 | 21.72 | 22.67 | 7.53 |
| *GIR* | 0.39 | 0.32 | 0.22 | 0.17 | $t(\alpha^*\text{-}GIR)$ | 33.42 | 25.27 | 24.58 | 13.20 |

Table 2. Cont'd

| | | | | | | | | | |
|---|---|---|---|---|---|---|---|---|---|
| ***Panel B: n = 10 funds*** | | | | | | | | | |
| | ***CAPM*** | ***FF3*** | ***Carhart4*** | ***FF5*** | | ***CAPM*** | ***FF3*** | ***Carhart4*** | ***FF5*** |
| ***B.1: L = 36 months*** | | | | | | | | | |
| $\alpha*$ | 0.82 | 0.52 | 0.42 | 0.27 | $t(\alpha*\text{-}IR)$ | 4.26 | 2.16 | 1.42 | 2.67 |
| $IR$ | 0.75 | 0.49 | 0.40 | 0.24 | $t(IR\text{-}GIR)$ | 2.72 | -2.65 | -3.42 | -0.87 |
| $GIR$ | 0.71 | 0.53 | 0.44 | 0.25 | $t(\alpha*\text{-}GIR)$ | 6.61 | -0.51 | -1.97 | 1.69 |
| ***B.2: L = 60 months*** | | | | | | | | | |
| $\alpha*$ | 0.61 | 0.40 | 0.31 | 0.22 | $t(\alpha*\text{-}IR)$ | 4.80 | 4.02 | 2.83 | 3.47 |
| $IR$ | 0.55 | 0.37 | 0.29 | 0.19 | $t(IR\text{-}GIR)$ | 6.77 | 1.66 | 0.85 | 1.87 |
| $GIR$ | 0.48 | 0.35 | 0.28 | 0.18 | $t(\alpha*\text{-}GIR)$ | 11.05 | 5.46 | 3.54 | 5.14 |
| ***B.3: L = 120 months*** | | | | | | | | | |
| $\alpha*$ | 0.47 | 0.36 | 0.25 | 0.18 | $t(\alpha*\text{-}IR)$ | 4.44 | 6.30 | 3.42 | 4.27 |
| $IR$ | 0.43 | 0.31 | 0.23 | 0.15 | $t(IR\text{-}GIR)$ | 8.17 | 4.55 | 4.40 | 2.95 |
| $GIR$ | 0.36 | 0.28 | 0.21 | 0.13 | $t(\alpha*\text{-}GIR)$ | 12.07 | 10.30 | 7.52 | 6.78 |
| ***B.4: L = 240 months*** | | | | | | | | | |
| $\alpha*$ | 0.42 | 0.36 | 0.24 | 0.19 | $t(\alpha*\text{-}IR)$ | 4.54 | 7.06 | 2.83 | 5.33 |
| $IR$ | 0.39 | 0.31 | 0.23 | 0.15 | $t(IR\text{-}GIR)$ | 9.10 | 7.05 | 8.27 | 3.27 |
| $GIR$ | 0.33 | 0.27 | 0.19 | 0.13 | $t(\alpha*\text{-}GIR)$ | 13.38 | 12.68 | 10.65 | 7.98 |
| ***B.5: L = 360 months*** | | | | | | | | | |
| $\alpha*$ | 0.40 | 0.34 | 0.22 | 0.18 | $t(\alpha*\text{-}IR)$ | 2.90 | 5.86 | 2.08 | 4.46 |
| $IR$ | 0.38 | 0.29 | 0.21 | 0.14 | $t(IR\text{-}GIR)$ | 9.07 | 7.09 | 9.82 | 2.57 |
| $GIR$ | 0.31 | 0.25 | 0.17 | 0.12 | $t(\alpha*\text{-}GIR)$ | 12.12 | 11.32 | 11.19 | 6.57 |
| ***Panel C: n = 50 funds*** | | | | | | | | | |
| | ***CAPM*** | ***FF3*** | ***Carhart4*** | ***FF5*** | | ***CAPM*** | ***FF3*** | ***Carhart4*** | ***FF5*** |
| ***C.1: L = 60 months*** | | | | | | | | | |
| $\alpha*$ | 1.45 | 0.96 | 0.74 | 0.52 | $t(\alpha*\text{-}IR)$ | 6.35 | 6.08 | 3.62 | 5.72 |
| $IR$ | 1.28 | 0.85 | 0.68 | 0.43 | $t(IR\text{-}GIR)$ | -22.03 | -21.10 | -18.67 | -16.27 |
| $GIR$ | 4.01 | 3.36 | 2.78 | 1.95 | $t(\alpha*\text{-}GIR)$ | -20.60 | -20.18 | -18.12 | -15.26 |
| ***C.2: L = 120 months*** | | | | | | | | | |
| $\alpha*$ | 1.10 | 0.85 | 0.59 | 0.45 | $t(\alpha*\text{-}IR)$ | 6.20 | 10.70 | 5.24 | 6.68 |
| $IR$ | 0.99 | 0.72 | 0.54 | 0.35 | $t(IR\text{-}GIR)$ | 26.30 | 19.97 | 15.89 | 10.14 |
| $GIR$ | 0.61 | 0.51 | 0.40 | 0.24 | $t(\alpha*\text{-}GIR)$ | 30.26 | 27.97 | 20.00 | 15.17 |
| ***C.3: L = 240 months*** | | | | | | | | | |
| $\alpha*$ | 0.97 | 0.83 | 0.55 | 0.46 | $t(\alpha*\text{-}IR)$ | 8.11 | 16.39 | 5.74 | 10.62 |
| $IR$ | 0.89 | 0.70 | 0.52 | 0.35 | $t(IR\text{-}GIR)$ | 46.75 | 40.10 | 34.80 | 15.52 |
| $GIR$ | 0.49 | 0.42 | 0.31 | 0.21 | $t(\alpha*\text{-}GIR)$ | 52.25 | 48.10 | 40.95 | 24.02 |
| ***C.4: L = 360 months*** | | | | | | | | | |
| $\alpha*$ | 0.91 | 0.78 | 0.51 | 0.45 | $t(\alpha*\text{-}IR)$ | 5.33 | 15.05 | 4.18 | 9.49 |
| $IR$ | 0.86 | 0.65 | 0.48 | 0.33 | $t(IR\text{-}GIR)$ | 46.68 | 39.98 | 35.29 | 13.49 |
| $GIR$ | 0.44 | 0.36 | 0.25 | 0.20 | $t(\alpha*\text{-}GIR)$ | 52.75 | 45.24 | 38.56 | 20.87 |

Table 2. Cont'd

| Panel D: n = 100 funds | | | | | | | | | |
|---|---|---|---|---|---|---|---|---|---|
| | ***CAPM*** | ***FF3*** | ***Carhart4*** | ***FF5*** | | ***CAPM*** | ***FF3*** | ***Carhart4*** | ***FF5*** |
| ***D.1: L = 120 months*** | | | | | | | | | |
| $\alpha^*$ | 1.56 | 1.22 | 0.84 | 0.65 | $t(\alpha^*\text{-}IR)$ | 6.49 | 12.40 | 5.53 | 7.33 |
| *IR* | 1.40 | 1.03 | 0.77 | 0.51 | $t(IR\text{-}GIR)$ | -5.85 | -8.06 | -8.83 | -6.34 |
| *GIR* | 1.60 | 1.27 | 1.00 | 0.68 | $t(\alpha^*\text{-}GIR)$ | -1.07 | -1.66 | -5.99 | -1.08 |
| ***D.2: L = 240 months*** | | | | | | | | | |
| $\alpha^*$ | 1.38 | 1.20 | 0.78 | 0.68 | $t(\alpha^*\text{-}IR)$ | 9.29 | 22.77 | 6.37 | 13.10 |
| *IR* | 1.26 | 0.99 | 0.73 | 0.50 | $t(IR\text{-}GIR)$ | 64.57 | 63.27 | 51.36 | 23.71 |
| *GIR* | 0.55 | 0.46 | 0.33 | 0.24 | $t(\alpha^*\text{-}GIR)$ | 71.36 | 75.18 | 59.08 | 34.23 |
| ***D.3: L = 360 months*** | | | | | | | | | |
| $\alpha^*$ | 1.29 | 1.13 | 0.72 | 0.66 | $t(\alpha^*\text{-}IR)$ | 5.97 | 22.84 | 4.87 | 11.95 |
| *IR* | 1.21 | 0.92 | 0.68 | 0.48 | $t(IR\text{-}GIR)$ | 64.10 | 67.88 | 50.37 | 22.19 |
| *GIR* | 0.47 | 0.39 | 0.27 | 0.22 | $t(\alpha^*\text{-}GIR)$ | 73.36 | 77.35 | 55.85 | 31.47 |

| Panel E: n = 200 funds | | | | | | | | | |
|---|---|---|---|---|---|---|---|---|---|
| | ***CAPM*** | ***FF3*** | ***FF4*** | ***FF5*** | | ***CAPM*** | ***FF3*** | ***FF4*** | ***FF5*** |
| ***E.1: L = 240 months*** | | | | | | | | | |
| $\alpha^*$ | 1.96 | 1.71 | 1.12 | 0.96 | $t(\alpha^*\text{-}IR)$ | 9.91 | 30.67 | 7.19 | 14.91 |
| *IR* | 1.78 | 1.41 | 1.04 | 0.71 | $t(IR\text{-}GIR)$ | 38.25 | 38.39 | 26.48 | 15.75 |
| *GIR* | 1.03 | 0.82 | 0.63 | 0.44 | $t(\alpha^*\text{-}GIR)$ | 45.72 | 55.72 | 32.16 | 27.87 |
| ***E.2: L = 360 months*** | | | | | | | | | |
| $\alpha^*$ | 1.83 | 1.61 | 1.02 | 0.94 | $t(\alpha^*\text{-}IR)$ | 6.37 | 36.23 | 5.34 | 13.75 |
| *IR* | 1.71 | 1.30 | 0.96 | 0.67 | $t(IR\text{-}GIR)$ | 81.71 | 115.61 | 69.92 | 32.64 |
| *GIR* | 0.49 | 0.39 | 0.27 | 0.22 | $t(\alpha^*\text{-}GIR)$ | 95.83 | 138.82 | 78.46 | 42.79 |

## Table 3. Average Distance Metrics: Monthly Performance

This table shows the average distance metrics for three performance measures ($\alpha^*$, *IR* and *GIR*) between a misspecified model and the true model. Four misspecified models (factors) are *CAPM* (*MKT*), *FF3* (*MKT SMB HML*), *Carhart4* (*MKT SMB HML UMD*), and *FF5* (*MKT SMB HML RMW CMA*). The true model is *FF6* (*MKT SMB HML RMW CMA UMD*). At each month starting from 1963:07, $n$ = 10, 25, 50, 100, 200 number of funds (shown in each sub-panel) are randomly selected from a universe of 346 Fama-French characteristics-sorted portfolios including 15×10 univariate-, 4×25 bivariate-, 3×32 triple-sorted portfolios. The selected funds are held for $L$ = 36, 60, 120, 240, and 360 months (shown in Panel A, B, C, D and E, respectively) for performance evaluation by $\alpha^*$, *IR* and *GIR* at the end of the period. Simulation is conducted on rolling window basis until the last month of the evaluation period reaches 2016:12 (the end of sample), creating overlapping time-series observations of distance metrics for $\alpha^*$, *IR* and *GIR* under each model. The values reported in the first four columns are the time-series average of the distance values under each model. All data are from the French Data Library.

The average distance metric is defined as the Euclidean 2-norm $||(\Sigma^{-\frac{1}{2}}\alpha - \Phi^{-\frac{1}{2}}a)/n||$ (i.e., square root of mean squared difference), where $\alpha$ and $a$ are alphas, $\Sigma$ and $\Phi$ are residual covariance matrices, produced by the misspecified model and the true model, respectively. To compute the distance metric, *GIR* uses full information in $\Sigma$ and $\Phi$, and $\Sigma^{-\frac{1}{2}}$ and $\Phi^{-\frac{1}{2}}$ are symmetric positive-definite inverse square root of $\Sigma$ and $\Phi$, respectively; *IR* uses only diagonal information in $\Sigma$ and $\Phi$ (i.e., $\Sigma^{-\frac{1}{2}}$ and $\Phi^{-\frac{1}{2}}$ are heteroskedastic diagonal matrices); and for $\alpha^*$, $\Sigma^{-\frac{1}{2}}$ and $\Phi^{-\frac{1}{2}}$ are homoscedastic diagonal matrices whose elements are set equal to the inverse of cross-sectional average standard error of residuals, respectively.

| | CAPM | FF3 | Carhart4 | FF5 | CAPM | FF3 | Carhart4 | FF5 | CAPM | FF3 | Carhart4 | FF5 |
|---|---|---|---|---|---|---|---|---|---|---|---|---|
| | | | | | **Panel A: L = 36 months** | | | | | | | |
| | | $n$ = 10 | | | | $n$ = 25 | | | | | | |
| $\alpha^*$ | 0.26 | 0.16 | 0.13 | 0.08 | 0.27 | 0.17 | 0.14 | 0.09 | | | | |
| IR | 0.24 | 0.16 | 0.13 | 0.07 | 0.24 | 0.16 | 0.13 | 0.08 | | | | |
| GIR | 0.22 | 0.17 | 0.14 | 0.08 | 0.50 | 0.41 | 0.35 | 0.21 | | | | |
| | | | | | **Panel B: L = 60 months** | | | | | | | |
| | | $n$ = 10 | | | | $n$ = 25 | | | | $n$ = 50 | | |
| $\alpha^*$ | 0.19 | 0.13 | 0.10 | 0.07 | 0.20 | 0.13 | 0.10 | 0.07 | 0.20 | 0.14 | 0.10 | 0.07 |
| IR | 0.17 | 0.12 | 0.09 | 0.06 | 0.18 | 0.12 | 0.09 | 0.06 | 0.18 | 0.12 | 0.10 | 0.06 |
| GIR | 0.15 | 0.11 | 0.09 | 0.06 | 0.15 | 0.12 | 0.10 | 0.06 | 0.57 | 0.48 | 0.39 | 0.28 |
| | | | | | **Panel C: L = 120 months** | | | | | | | |
| | | $n$ = 10 | | | | $n$ = 25 | | | | $n$ = 50 | | |
| $\alpha^*$ | 0.15 | 0.11 | 0.08 | 0.06 | 0.15 | 0.12 | 0.08 | 0.06 | 0.16 | 0.12 | 0.08 | 0.06 |
| IR | 0.14 | 0.10 | 0.07 | 0.05 | 0.14 | 0.10 | 0.08 | 0.05 | 0.14 | 0.10 | 0.08 | 0.05 |
| GIR | 0.11 | 0.09 | 0.07 | 0.04 | 0.10 | 0.08 | 0.06 | 0.04 | 0.09 | 0.07 | 0.06 | 0.03 |
| | | $n$ = 100 | | | | | | | | | | |
| $\alpha^*$ | 0.16 | 0.12 | 0.08 | 0.06 | | | | | | | | |
| IR | 0.14 | 0.10 | 0.08 | 0.05 | | | | | | | | |
| GIR | 0.16 | 0.13 | 0.10 | 0.07 | | | | | | | | |
| | | | | | **Panel D: L = 240 months** | | | | | | | |
| | | $n$ = 10 | | | | $n$ = 25 | | | | $n$ = 50 | | |
| $\alpha^*$ | 0.13 | 0.11 | 0.08 | 0.06 | 0.14 | 0.12 | 0.08 | 0.06 | 0.14 | 0.12 | 0.08 | 0.07 |
| IR | 0.12 | 0.10 | 0.07 | 0.05 | 0.13 | 0.10 | 0.07 | 0.05 | 0.13 | 0.10 | 0.07 | 0.05 |
| GIR | 0.11 | 0.08 | 0.06 | 0.04 | 0.09 | 0.07 | 0.05 | 0.04 | 0.07 | 0.06 | 0.04 | 0.03 |
| | | $n$ = 100 | | | | $n$ = 200 | | | | | | |
| $\alpha^*$ | 0.14 | 0.12 | 0.08 | 0.07 | 0.14 | 0.12 | 0.08 | 0.07 | | | | |
| IR | 0.13 | 0.10 | 0.07 | 0.05 | 0.13 | 0.10 | 0.07 | 0.05 | | | | |
| GIR | 0.05 | 0.05 | 0.03 | 0.02 | 0.07 | 0.06 | 0.04 | 0.03 | | | | |
| | | | | | **Panel E: L = 360 months** | | | | | | | |
| | | $n$ = 10 | | | | $n$ = 25 | | | | $n$ = 50 | | |
| $\alpha^*$ | 0.13 | 0.11 | 0.07 | 0.06 | 0.13 | 0.11 | 0.07 | 0.06 | 0.13 | 0.11 | 0.07 | 0.06 |
| IR | 0.12 | 0.09 | 0.07 | 0.04 | 0.12 | 0.09 | 0.07 | 0.05 | 0.12 | 0.09 | 0.07 | 0.05 |
| GIR | 0.10 | 0.08 | 0.05 | 0.04 | 0.08 | 0.06 | 0.04 | 0.03 | 0.06 | 0.05 | 0.04 | 0.03 |
| | | $n$ = 100 | | | | $n$ = 200 | | | | | | |
| $\alpha^*$ | 0.13 | 0.11 | 0.07 | 0.07 | 0.13 | 0.11 | 0.07 | 0.07 | | | | |
| IR | 0.12 | 0.09 | 0.07 | 0.05 | 0.12 | 0.09 | 0.07 | 0.05 | | | | |
| GIR | 0.05 | 0.04 | 0.03 | 0.02 | 0.03 | 0.03 | 0.02 | 0.02 | | | | |

Table 4. Monthly Out-of-Sample Performance of Long-Short Fund Portfolios

This table reports the time-series average alpha $A(\alpha)$, alpha standard deviation $\sigma(\alpha)$, $t$-statistic $t(\alpha)$ adjusted for serial correlations, and cross-model alpha standard deviation $std(\alpha)$ for long-short portfolios formed below. At each month starting from 1963:07, $n$ = 10, 25, 50, 100, 200 number of funds (shown in each sub-panel) are randomly selected from a universe of 346 Fama-French characteristics-sorted portfolios including 15×10 univariate-, 4×25 bivariate-, 3×32 triple-sorted portfolios. The selected funds are held for $L$ = 36, 60, 120, 240, and 360 months (shown in Panel A, B, C, D and E, respectively) for ranking by $\alpha^*$, $IR$ and $GIR$ at the end of the period. A portfolio is formed by taking long positions on top quintile funds and short positions on bottom quintile funds ranked by three performance measures ($\alpha^*$, $IR$ and $GIR$) under each model (*CAPM FF3 Carhart4 FF5*). The portfolio is held for 12 months and alpha is computed from time-series regression over this out-of-sample period. Simulation is conducted on rolling window basis until the last month of the out-of-sample period reaches 2016:12 (the end of sample), creating overlapping time-series observations of portfolio returns. Performance measures are defined as: $GIR = \Sigma^{-1/2}\alpha$, $IR = \mathrm{D}^{-1/2}\alpha$, $\alpha^* = \alpha/\sigma_e$, where $\Sigma^{-\frac{1}{2}}$ is a symmetric positive-definite matrix as the inverse square root of $\Sigma$, $D$ uses only diagonal elements of $\Sigma$, and $\sigma_e$ is the cross-sectional average of standard errors of residuals. All data are from the French Data Library.

| | **CAPM** | | | **FF3** | | | **Carhart4** | | | **FF5** | | | **std ($\alpha$)** | | |
|---|---|---|---|---|---|---|---|---|---|---|---|---|---|---|---|
| | $\alpha$* | IR | GIR | $\alpha$* | IR | GIR | $\alpha$* | IR | GIR | $\alpha$* | IR | GIR | $\alpha$* | IR | GIR |
| | **Panel A: L = 36 months** | | | | | | | | | | | | | | |
| | $n$ = 10 funds | | | | | | | | | | | | | | |
| $A(\alpha)$ | 0.28 | 0.22 | 0.17 | 0.13 | 0.14 | 0.14 | 0.14 | 0.13 | 0.15 | 0.09 | 0.12 | 0.08 | 0.072 | 0.040 | 0.036 |
| $\sigma(\alpha)$ | 1.10 | 0.99 | 0.82 | 0.90 | 0.75 | 0.75 | 0.73 | 0.71 | 0.69 | 1.00 | 0.91 | 0.84 | | | |
| $t(\alpha)$ | 3.71 | 3.38 | 3.91 | 2.70 | 3.49 | 3.51 | 3.74 | 3.46 | 4.35 | 1.85 | 2.48 | 1.91 | | | |
| | $n$ = 25 funds | | | | | | | | | | | | | | |
| $A(\alpha)$ | 0.32 | 0.25 | 0.12 | 0.16 | 0.17 | 0.11 | 0.16 | 0.18 | 0.14 | 0.07 | 0.06 | 0.07 | 0.091 | 0.065 | 0.023 |
| $\sigma(\alpha)$ | 0.95 | 0.84 | 0.51 | 0.62 | 0.56 | 0.46 | 0.56 | 0.53 | 0.44 | 0.71 | 0.64 | 0.48 | | | |
| $t(\alpha)$ | 4.17 | 3.77 | 3.95 | 3.99 | 4.70 | 4.16 | 4.62 | 5.40 | 6.13 | 1.62 | 1.80 | 3.14 | | | |
| | **Panel B: L = 60 months** | | | | | | | | | | | | | | |
| | $n$ = 10 funds | | | | | | | | | | | | | | |
| $A(\alpha)$ | 0.22 | 0.20 | 0.25 | 0.28 | 0.28 | 0.24 | 0.25 | 0.25 | 0.22 | 0.08 | 0.09 | 0.07 | 0.077 | 0.073 | 0.072 |
| $\sigma(\alpha)$ | 0.98 | 0.91 | 0.80 | 0.76 | 0.72 | 0.67 | 0.79 | 0.73 | 0.70 | 0.89 | 0.87 | 0.82 | | | |
| $t(\alpha)$ | 3.54 | 3.40 | 5.29 | 6.92 | 7.26 | 6.65 | 5.64 | 6.13 | 5.56 | 1.98 | 2.29 | 2.11 | | | |
| | $n$ = 25 funds | | | | | | | | | | | | | | |
| $A(\alpha)$ | 0.24 | 0.22 | 0.22 | 0.25 | 0.25 | 0.20 | 0.24 | 0.23 | 0.19 | 0.12 | 0.12 | 0.06 | 0.052 | 0.049 | 0.060 |
| $\sigma(\alpha)$ | 0.81 | 0.74 | 0.50 | 0.54 | 0.51 | 0.45 | 0.54 | 0.50 | 0.45 | 0.61 | 0.56 | 0.51 | | | |
| $t(\alpha)$ | 3.82 | 3.74 | 6.36 | 7.07 | 7.52 | 7.29 | 7.01 | 7.29 | 6.92 | 3.68 | 3.97 | 2.42 | | | |
| | $n$ = 50 funds | | | | | | | | | | | | | | |
| $A(\alpha)$ | 0.26 | 0.23 | 0.14 | 0.24 | 0.24 | 0.11 | 0.24 | 0.25 | 0.09 | 0.12 | 0.12 | 0.05 | 0.057 | 0.053 | 0.034 |
| $\sigma(\alpha)$ | 0.74 | 0.67 | 0.32 | 0.46 | 0.45 | 0.29 | 0.42 | 0.40 | 0.27 | 0.49 | 0.46 | 0.31 | | | |
| $t(\alpha)$ | 4.07 | 4.00 | 6.77 | 6.94 | 7.12 | 6.32 | 8.01 | 8.67 | 6.18 | 3.73 | 3.96 | 3.44 | | | |
| | **Panel C: L = 120 months** | | | | | | | | | | | | | | |
| | $n$ = 10 funds | | | | | | | | | | | | | | |
| $A(\alpha)$ | 0.41 | 0.37 | 0.30 | 0.24 | 0.22 | 0.22 | 0.25 | 0.25 | 0.26 | 0.07 | 0.05 | 0.09 | 0.120 | 0.116 | 0.080 |
| $\sigma(\alpha)$ | 0.94 | 0.88 | 0.83 | 0.80 | 0.76 | 0.71 | 0.72 | 0.73 | 0.70 | 0.88 | 0.83 | 0.80 | | | |
| $t(\alpha)$ | 6.65 | 6.95 | 6.00 | 5.29 | 5.17 | 5.60 | 6.10 | 6.36 | 6.79 | 1.44 | 1.23 | 2.27 | | | |
| | $n$ = 25 funds | | | | | | | | | | | | | | |
| $A(\alpha)$ | 0.45 | 0.40 | 0.31 | 0.24 | 0.24 | 0.22 | 0.27 | 0.26 | 0.23 | 0.06 | 0.09 | 0.09 | 0.137 | 0.110 | 0.076 |
| $\sigma(\alpha)$ | 0.79 | 0.74 | 0.53 | 0.63 | 0.58 | 0.49 | 0.57 | 0.53 | 0.49 | 0.68 | 0.61 | 0.57 | | | |
| $t(\alpha)$ | 7.46 | 7.51 | 8.46 | 5.99 | 5.97 | 6.75 | 6.86 | 6.91 | 7.46 | 1.56 | 2.60 | 3.16 | | | |
| | $n$ = 50 funds | | | | | | | | | | | | | | |
| $A(\alpha)$ | 0.43 | 0.39 | 0.25 | 0.22 | 0.21 | 0.19 | 0.26 | 0.24 | 0.20 | 0.08 | 0.09 | 0.11 | 0.124 | 0.107 | 0.049 |
| $\sigma(\alpha)$ | 0.71 | 0.62 | 0.36 | 0.50 | 0.49 | 0.34 | 0.47 | 0.45 | 0.37 | 0.53 | 0.47 | 0.38 | | | |
| $t(\alpha)$ | 7.20 | 7.72 | 9.31 | 5.46 | 5.24 | 7.92 | 6.65 | 6.72 | 7.23 | 2.28 | 2.81 | 5.16 | | | |
| | $n$ = 100 funds | | | | | | | | | | | | | | |
| $A(\alpha)$ | 0.45 | 0.40 | 0.16 | 0.25 | 0.24 | 0.13 | 0.27 | 0.26 | 0.12 | 0.09 | 0.10 | 0.08 | 0.128 | 0.104 | 0.027 |
| $\sigma(\alpha)$ | 0.65 | 0.57 | 0.22 | 0.44 | 0.43 | 0.20 | 0.40 | 0.39 | 0.21 | 0.45 | 0.42 | 0.23 | | | |
| $t(\alpha)$ | 7.54 | 7.71 | 9.46 | 6.26 | 6.21 | 8.72 | 7.56 | 7.57 | 7.87 | 2.54 | 3.13 | 6.30 | | | |

Table 4. Cont'd

| | CAPM | | | FF3 | | | Carhart4 | | | FF5 | | | std($\alpha$) | | |
|---|---|---|---|---|---|---|---|---|---|---|---|---|---|---|---|
| | **Panel D: L = 240 months** | | | | | | | | | | | | | | |
| | $n$ = 10 funds | | | | | | | | | | | | | | |
| $A(\alpha)$ | 0.23 | 0.22 | 0.30 | 0.26 | 0.27 | 0.29 | 0.33 | 0.33 | 0.34 | 0.13 | 0.12 | 0.14 | 0.072 | 0.076 | 0.075 |
| $\sigma(\alpha)$ | 0.99 | 0.94 | 0.81 | 0.82 | 0.79 | 0.72 | 0.74 | 0.74 | 0.68 | 0.93 | 0.87 | 0.84 | | | |
| $t(\alpha)$ | 3.05 | 2.99 | 4.83 | 4.64 | 5.16 | 5.69 | 6.85 | 6.44 | 7.56 | 2.48 | 2.53 | 3.11 | | | |
| | $n$ = 25 funds | | | | | | | | | | | | | | |
| $A(\alpha)$ | 0.34 | 0.31 | 0.37 | 0.32 | 0.32 | 0.28 | 0.35 | 0.36 | 0.30 | 0.12 | 0.12 | 0.13 | 0.094 | 0.092 | 0.088 |
| $\sigma(\alpha)$ | 0.81 | 0.76 | 0.57 | 0.54 | 0.53 | 0.48 | 0.53 | 0.54 | 0.47 | 0.63 | 0.58 | 0.53 | | | |
| $t(\alpha)$ | 4.54 | 4.75 | 7.46 | 7.00 | 7.25 | 7.08 | 8.17 | 8.11 | 8.02 | 3.08 | 3.33 | 3.81 | | | |
| | $n$ = 50 funds | | | | | | | | | | | | | | |
| $A(\alpha)$ | 0.34 | 0.31 | 0.32 | 0.27 | 0.28 | 0.25 | 0.33 | 0.34 | 0.25 | 0.11 | 0.12 | 0.09 | 0.090 | 0.087 | 0.082 |
| $\sigma(\alpha)$ | 0.72 | 0.67 | 0.42 | 0.50 | 0.49 | 0.39 | 0.49 | 0.46 | 0.36 | 0.50 | 0.45 | 0.36 | | | |
| $t(\alpha)$ | 4.62 | 4.65 | 7.85 | 5.61 | 6.01 | 7.04 | 7.16 | 7.74 | 7.61 | 3.14 | 3.69 | 3.96 | | | |
| | $n$ = 100 funds | | | | | | | | | | | | | | |
| $A(\alpha)$ | 0.34 | 0.33 | 0.26 | 0.29 | 0.29 | 0.20 | 0.33 | 0.33 | 0.21 | 0.12 | 0.12 | 0.13 | 0.090 | 0.086 | 0.050 |
| $\sigma(\alpha)$ | 0.68 | 0.63 | 0.29 | 0.46 | 0.44 | 0.28 | 0.41 | 0.41 | 0.26 | 0.41 | 0.38 | 0.27 | | | |
| $t(\alpha)$ | 4.58 | 4.76 | 9.05 | 6.00 | 6.52 | 7.91 | 7.79 | 7.99 | 8.45 | 3.11 | 3.64 | 6.57 | | | |
| | $n$ = 200 funds | | | | | | | | | | | | | | |
| $A(\alpha)$ | 0.36 | 0.34 | 0.17 | 0.29 | 0.29 | 0.14 | 0.33 | 0.33 | 0.13 | 0.12 | 0.13 | 0.09 | 0.090 | 0.087 | 0.029 |
| $\sigma(\alpha)$ | 0.66 | 0.60 | 0.17 | 0.43 | 0.42 | 0.16 | 0.38 | 0.37 | 0.15 | 0.38 | 0.34 | 0.15 | | | |
| $t(\alpha)$ | 4.77 | 5.07 | 10.83 | 6.14 | 6.37 | 8.79 | 7.85 | 8.20 | 9.05 | 3.23 | 3.67 | 7.59 | | | |
| | **Panel E: L = 360 months** | | | | | | | | | | | | | | |
| | $n$ = 10 funds | | | | | | | | | | | | | | |
| $A(\alpha)$ | 0.21 | 0.21 | 0.26 | 0.12 | 0.14 | 0.18 | 0.23 | 0.24 | 0.23 | -0.06 | -0.03 | 0.00 | 0.113 | 0.107 | 0.102 |
| $\sigma(\alpha)$ | 1.10 | 1.06 | 0.97 | 0.93 | 0.93 | 0.85 | 0.86 | 0.88 | 0.85 | 1.11 | 1.03 | 1.00 | | | |
| $t(\alpha)$ | 1.98 | 2.02 | 2.92 | 1.58 | 1.82 | 2.83 | 3.60 | 3.56 | 3.55 | -0.70 | -0.47 | -0.04 | | | |
| | $n$ = 25 funds | | | | | | | | | | | | | | |
| $A(\alpha)$ | 0.35 | 0.33 | 0.34 | 0.23 | 0.22 | 0.22 | 0.28 | 0.27 | 0.26 | 0.03 | 0.03 | 0.10 | 0.118 | 0.110 | 0.086 |
| $\sigma(\alpha)$ | 0.93 | 0.89 | 0.73 | 0.64 | 0.64 | 0.54 | 0.60 | 0.58 | 0.55 | 0.69 | 0.64 | 0.58 | | | |
| $t(\alpha)$ | 3.30 | 3.28 | 4.28 | 3.45 | 3.41 | 4.32 | 5.04 | 4.91 | 4.88 | 0.51 | 0.63 | 2.08 | | | |
| | $n$ = 50 funds | | | | | | | | | | | | | | |
| $A(\alpha)$ | 0.37 | 0.35 | 0.30 | 0.24 | 0.25 | 0.20 | 0.29 | 0.28 | 0.25 | 0.07 | 0.07 | 0.12 | 0.108 | 0.105 | 0.064 |
| $\sigma(\alpha)$ | 0.80 | 0.75 | 0.51 | 0.55 | 0.53 | 0.41 | 0.50 | 0.46 | 0.41 | 0.56 | 0.50 | 0.42 | | | |
| $t(\alpha)$ | 3.72 | 3.85 | 5.10 | 3.92 | 4.23 | 4.74 | 5.60 | 5.79 | 5.93 | 1.32 | 1.33 | 3.04 | | | |
| | $n$ = 100 funds | | | | | | | | | | | | | | |
| $A(\alpha)$ | 0.37 | 0.37 | 0.27 | 0.25 | 0.23 | 0.19 | 0.29 | 0.29 | 0.22 | 0.10 | 0.09 | 0.11 | 0.099 | 0.099 | 0.056 |
| $\sigma(\alpha)$ | 0.77 | 0.72 | 0.38 | 0.51 | 0.47 | 0.30 | 0.41 | 0.39 | 0.28 | 0.46 | 0.42 | 0.30 | | | |
| $t(\alpha)$ | 3.67 | 3.89 | 5.85 | 3.99 | 3.97 | 5.49 | 5.93 | 6.00 | 6.93 | 2.06 | 2.12 | 3.54 | | | |
| | $n$ = 200 funds | | | | | | | | | | | | | | |
| $A(\alpha)$ | 0.38 | 0.37 | 0.21 | 0.24 | 0.23 | 0.15 | 0.29 | 0.28 | 0.16 | 0.09 | 0.09 | 0.10 | 0.104 | 0.101 | 0.038 |
| $\sigma(\alpha)$ | 0.77 | 0.72 | 0.26 | 0.50 | 0.47 | 0.24 | 0.37 | 0.37 | 0.20 | 0.44 | 0.39 | 0.24 | | | |
| $t(\alpha)$ | 3.57 | 3.79 | 6.24 | 3.73 | 3.66 | 4.89 | 5.93 | 5.80 | 6.55 | 1.69 | 1.97 | 3.41 | | | |

Table 5. Weekly Statistics for Distance Metrics and Mean Difference Test

This table reports summary statistics of total distance (*TD*) in the first four columns, average distance (*AD*) in the middle four columns, and mean difference *t*-statistics in the last four columns using weekly data. For each performance measure ($\alpha^*$, *IR* and *GIR*), *TD* and *AD* are calculated between a misspecified model and the true model. Four misspecified models (factors) are *CAPM* (*MKT*), *FF3* (*MKT SMB HML*), *Carhart4* (*MKT SMB HML UMD*), and *FF5* (*MKT SMB HML RMW CMA*). The true model is *FF6* (*MKT SMB HML RMW CMA UMD*). At each week starting from July 1963, $n$ = 10, 25, 50, 75, 100 number of funds (shown in Panel A, B, C, D, and E, respectively) are randomly selected from a universe of 175 Fama-French characteristics-sorted portfolios including 25 *Size-B/M* portfolios, 25 *Size-OP* portfolios, 25 *Size-INV* portfolios, 25 *B/M-OP* portfolios, 25 *B/M-INV* portfolios, 25 *OP-INV* portfolios, and 25 *Size-MOM* portfolios. Daily returns are cumulated to compute weekly returns for these 175 portfolios and also for the six benchmark factors. The selected funds are held for $L$ = 156 weeks (36 months) for performance evaluation by the distance metrics at the end of the period.

Total distance (*TD*) is defined as the Euclidean 2-norm $||(\Sigma^{-\frac{1}{2}}\alpha - \Phi^{-\frac{1}{2}}a)||$ (i.e., square root of sum squared difference), and average distance (*AD*) is defined as $||(\Sigma^{-\frac{1}{2}}\alpha - \Phi^{-\frac{1}{2}}a)/n||$ (i.e., square root of mean squared difference). $\alpha$ and $a$ are alphas, $\Sigma$ and $\Phi$ are residual covariance matrices, produced by the misspecified model and the true model, respectively. To compute the distance metric, *GIR* uses full information in $\Sigma$ and $\Phi$, and $\Sigma^{-\frac{1}{2}}$ and $\Phi^{-\frac{1}{2}}$ are symmetric positive-definite inverse square root of $\Sigma$ and $\Phi$, respectively; *IR* uses only diagonal information in $\Sigma$ and $\Phi$ (i.e., $\Sigma^{-\frac{1}{2}}$ and $\Phi^{-\frac{1}{2}}$ are heteroskedastic diagonal matrices); and for $\alpha^*$, $\Sigma^{-\frac{1}{2}}$ and $\Phi^{-\frac{1}{2}}$ are homoscedastic diagonal matrices whose elements are set to the inverse of cross-sectional average standard error of residuals, respectively.

| | *CAPM* | *FF3* | *Carhart4* | *FF5* | | *CAPM* | *FF3* | *Carhart4* | *FF5* | | *CAPM* | *FF3* | *Carhart4* | *FF5* |
|---|---|---|---|---|---|---|---|---|---|---|---|---|---|---|
| | | | | | | | ***Panel A: n = 10 funds*** | | | | | | | |
| | *Total Distance (TD)* | | | | | *Average Distance (AD)* | | | | | *Mean Difference Test* | | | |
| $\alpha^*$ | 0.34 | 0.20 | 0.16 | 0.09 | $\alpha^*$ | 0.108 | 0.062 | 0.052 | 0.028 | $t(\alpha^*\text{-}IR)$ | 4.78 | 7.64 | 5.90 | 4.90 |
| *IR* | 0.32 | 0.18 | 0.15 | 0.08 | *IR* | 0.102 | 0.056 | 0.047 | 0.026 | $t(IR\text{-}GIR)$ | 22.33 | 7.48 | 5.09 | 6.42 |
| *GIR* | 0.25 | 0.16 | 0.14 | 0.07 | *GIR* | 0.079 | 0.051 | 0.044 | 0.022 | $t(\alpha^*\text{-}GIR)$ | 25.61 | 14.67 | 10.89 | 10.58 |
| | | | | | | | ***Panel B: n = 25 funds*** | | | | | | | |
| | *Total Distance (TD)* | | | | | *Average Distance (AD)* | | | | | *Mean Difference Test* | | | |
| $\alpha^*$ | 0.55 | 0.32 | 0.27 | 0.15 | $\alpha^*$ | 0.111 | 0.064 | 0.053 | 0.030 | $t(\alpha^*\text{-}IR)$ | 6.13 | 10.16 | 7.43 | 6.80 |
| *IR* | 0.52 | 0.28 | 0.24 | 0.13 | *IR* | 0.103 | 0.057 | 0.048 | 0.027 | $t(IR\text{-}GIR)$ | 41.78 | 17.79 | 12.56 | 14.78 |
| *GIR* | 0.33 | 0.23 | 0.20 | 0.10 | *GIR* | 0.066 | 0.047 | 0.040 | 0.020 | $t(\alpha^*\text{-}GIR)$ | 44.61 | 26.77 | 19.52 | 19.38 |
| | | | | | | | ***Panel C: n = 50 funds*** | | | | | | | |
| | *Total Distance (TD)* | | | | | *Average Distance (AD)* | | | | | *Mean Difference Test* | | | |
| $\alpha^*$ | 0.79 | 0.45 | 0.38 | 0.22 | $\alpha^*$ | 0.112 | 0.064 | 0.054 | 0.031 | $t(\alpha^*\text{-}IR)$ | 6.65 | 11.34 | 8.01 | 7.68 |
| *IR* | 0.73 | 0.40 | 0.34 | 0.19 | *IR* | 0.104 | 0.057 | 0.048 | 0.027 | $t(IR\text{-}GIR)$ | 54.31 | 25.72 | 18.26 | 21.24 |
| *GIR* | 0.41 | 0.31 | 0.27 | 0.13 | *GIR* | 0.058 | 0.044 | 0.038 | 0.019 | $t(\alpha^*\text{-}GIR)$ | 56.19 | 35.55 | 25.53 | 25.48 |
| | | | | | | | ***Panel D: n = 75 funds*** | | | | | | | |
| | *Total Distance (TD)* | | | | | *Average Distance (AD)* | | | | | *Mean Difference Test* | | | |
| $\alpha^*$ | 0.97 | 0.56 | 0.47 | 0.27 | $\alpha^*$ | 0.112 | 0.065 | 0.054 | 0.031 | $t(\alpha^*\text{-}IR)$ | 6.83 | 11.75 | 8.27 | 8.02 |
| *IR* | 0.90 | 0.50 | 0.42 | 0.24 | *IR* | 0.104 | 0.057 | 0.048 | 0.027 | $t(IR\text{-}GIR)$ | 58.32 | 26.62 | 18.88 | 22.39 |
| *GIR* | 0.48 | 0.38 | 0.33 | 0.16 | *GIR* | 0.056 | 0.043 | 0.038 | 0.019 | $t(\alpha^*\text{-}GIR)$ | 59.88 | 36.90 | 26.45 | 26.56 |
| | | | | | | | ***Panel E: n = 100 funds*** | | | | | | | |
| | *Total Distance (TD)* | | | | | *Average Distance (AD)* | | | | | *Mean Difference Test* | | | |
| $\alpha^*$ | 1.12 | 0.65 | 0.54 | 0.31 | $\alpha^*$ | 0.112 | 0.065 | 0.054 | 0.031 | $t(\alpha^*\text{-}IR)$ | 6.86 | 11.82 | 8.29 | 8.07 |
| *IR* | 1.04 | 0.57 | 0.49 | 0.27 | *IR* | 0.104 | 0.057 | 0.049 | 0.027 | $t(IR\text{-}GIR)$ | 53.65 | 18.17 | 12.69 | 16.22 |
| *GIR* | 0.59 | 0.47 | 0.41 | 0.21 | *GIR* | 0.059 | 0.047 | 0.041 | 0.021 | $t(\alpha^*\text{-}GIR)$ | 55.89 | 29.11 | 20.72 | 21.53 |

Table 6. Weekly Out-of-Sample Performance of Long-Short Fund Portfolios

This table reports the time-series average alpha $A(\alpha)$, alpha standard deviation $\sigma(\alpha)$, $t$-statistic $t(\alpha)$ adjusted for serial correlations, and cross-model alpha standard deviation $std(\alpha)$ for long-short portfolios formed below. At each week starting from July 1963, $n$ = 10, 25, 50, 75, 100 number of funds (shown in Panel A, B, C, D, and E, respectively) are randomly selected from a universe of 175 Fama-French characteristics-sorted portfolios including 25 *Size-B/M* portfolios, 25 *Size-OP* portfolios, 25 *Size-INV* portfolios, 25 *B/M-OP* portfolios, 25 *B/M-INV* portfolios, 25 *OP-INV* portfolios, and 25 *Size-MOM* portfolios. Daily returns are cumulated to compute weekly returns for these 175 portfolios and also for the six benchmark factors. The selected funds are held for $L$ = 156 weeks (36 months) for ranking by $\alpha^*$, *IR* and *GIR* at the end of the holding period. A portfolio is formed by taking long positions on top quintile funds and short positions on bottom quintile funds ranked by $\alpha^*$, *IR* and *GIR* under each model (*CAPM FF3 Carhart4 FF5*). The portfolio is held for 26 weeks (6 months) and alpha is computed from time-series regression over this out-of-sample period. Simulation is conducted on rolling window basis until the last week of the out-of-sample period reaches the end of sample, creating overlapping time-series observations of portfolio returns. Performance measures are defined as: $GIR = \Sigma^{-1/2}\alpha$, $IR = \mathrm{D}^{-1/2}\alpha$, $\alpha^* = \alpha/\sigma_e$, where $\Sigma^{-\frac{1}{2}}$ is a symmetric positive-definite matrix as the inverse square root of Σ, $D$ uses only diagonal elements of Σ, $D$ uses only diagonal elements of Σ, and $\sigma_e$ is the cross-sectional average of standard errors of residuals. All data are from the French Data Library.

| | CAPM | | | FF3 | | | Carhart4 | | | FF5 | | | std(α) | | |
|---|---|---|---|---|---|---|---|---|---|---|---|---|---|---|---|
| **Panel A: n = 10 funds** | | | | | | | | | | | | | | | |
| | $\alpha^*$ | IR | GIR | $\alpha^*$ | IR | GIR | $\alpha^*$ | IR | GIR | $\alpha^*$ | IR | GIR | $\alpha^*$ | IR | GIR |
| $A(\alpha)$ | 0.08 | 0.08 | 0.07 | 0.05 | 0.06 | 0.05 | 0.05 | 0.05 | 0.05 | 0.04 | 0.04 | 0.04 | 0.016 | 0.012 | 0.009 |
| $\sigma(\alpha)$ | 0.56 | 0.54 | 0.42 | 0.36 | 0.34 | 0.32 | 0.30 | 0.29 | 0.28 | 0.30 | 0.28 | 0.28 | | | |
| $t(\alpha)$ | 7.28 | 7.22 | 8.14 | 7.16 | 8.37 | 8.53 | 7.90 | 8.66 | 8.76 | 6.53 | 7.93 | 7.69 | | | |
| **Panel B: n = 25 funds** | | | | | | | | | | | | | | | |
| | $\alpha^*$ | IR | GIR | $\alpha^*$ | IR | GIR | $\alpha^*$ | IR | GIR | $\alpha^*$ | IR | GIR | $\alpha^*$ | IR | GIR |
| $A(\alpha)$ | 0.08 | 0.07 | 0.06 | 0.05 | 0.06 | 0.05 | 0.05 | 0.05 | 0.05 | 0.04 | 0.04 | 0.04 | 0.015 | 0.011 | 0.006 |
| $\sigma(\alpha)$ | 0.56 | 0.54 | 0.31 | 0.32 | 0.30 | 0.25 | 0.25 | 0.24 | 0.21 | 0.25 | 0.24 | 0.21 | | | |
| $t(\alpha)$ | 7.36 | 6.77 | 9.21 | 8.49 | 9.49 | 11.01 | 9.70 | 10.48 | 12.08 | 7.83 | 9.06 | 10.05 | | | |
| **Panel C: n = 50 funds** | | | | | | | | | | | | | | | |
| | $\alpha^*$ | IR | GIR | $\alpha^*$ | IR | GIR | $\alpha^*$ | IR | GIR | $\alpha^*$ | IR | GIR | $\alpha^*$ | IR | GIR |
| $A(\alpha)$ | 0.08 | 0.07 | 0.04 | 0.05 | 0.05 | 0.04 | 0.04 | 0.05 | 0.04 | 0.04 | 0.04 | 0.03 | 0.016 | 0.011 | 0.004 |
| $\sigma(\alpha)$ | 0.56 | 0.54 | 0.24 | 0.29 | 0.28 | 0.20 | 0.23 | 0.22 | 0.17 | 0.23 | 0.22 | 0.17 | | | |
| $t(\alpha)$ | 7.28 | 6.70 | 9.31 | 8.63 | 9.78 | 11.49 | 10.09 | 11.17 | 12.83 | 8.37 | 9.39 | 10.53 | | | |
| **Panel D: n = 75 funds** | | | | | | | | | | | | | | | |
| | $\alpha^*$ | IR | GIR | $\alpha^*$ | IR | GIR | $\alpha^*$ | IR | GIR | $\alpha^*$ | IR | GIR | $\alpha^*$ | IR | GIR |
| $A(\alpha)$ | 0.08 | 0.07 | 0.03 | 0.05 | 0.05 | 0.04 | 0.04 | 0.05 | 0.04 | 0.04 | 0.04 | 0.03 | 0.016 | 0.011 | 0.004 |
| $\sigma(\alpha)$ | 0.56 | 0.54 | 0.20 | 0.29 | 0.28 | 0.17 | 0.22 | 0.21 | 0.15 | 0.22 | 0.21 | 0.14 | | | |
| $t(\alpha)$ | 7.13 | 6.70 | 8.80 | 8.72 | 9.92 | 12.59 | 10.07 | 11.24 | 13.55 | 8.45 | 9.87 | 11.79 | | | |
| **Panel E: n = 100 funds** | | | | | | | | | | | | | | | |
| | $\alpha^*$ | IR | GIR | $\alpha^*$ | IR | GIR | $\alpha^*$ | IR | GIR | $\alpha^*$ | IR | GIR | $\alpha^*$ | IR | GIR |
| $A(\alpha)$ | 0.08 | 0.07 | 0.03 | 0.05 | 0.05 | 0.04 | 0.05 | 0.05 | 0.03 | 0.04 | 0.04 | 0.03 | 0.016 | 0.011 | 0.003 |
| $\sigma(\alpha)$ | 0.56 | 0.54 | 0.16 | 0.29 | 0.28 | 0.14 | 0.21 | 0.21 | 0.12 | 0.22 | 0.21 | 0.12 | | | |
| $t(\alpha)$ | 7.27 | 6.70 | 9.39 | 9.08 | 10.19 | 13.33 | 10.83 | 11.84 | 14.30 | 9.00 | 10.59 | 12.51 | | | |